\documentstyle [12pt,eqsecnum,aps] {revtex}
\tighten
\draft
\input epsf
\newcommand{\gtwid}{\mathrel{\raise.3ex\hbox{$>$\kern-.75em\lower1ex
\hbox{$\sim$}}}}
\newcommand{\ltwid}{\mathrel{\raise.3ex\hbox{$<$\kern-.75em\lower1ex
\hbox{$\sim$}}}}
\newcommand{\be}{\begin{equation}}
\newcommand{\ee}{\end{equation}}
\newcommand{\ba}{\begin{eqnarray}}
\newcommand{\ea}{\end{eqnarray}}

\newcommand{\beq}{\begin{equation}}
\newcommand{\eeq}{\end{equation}}
\newcommand{\beqs}{\begin{eqnarray}}
\newcommand{\eeqs}{\end{eqnarray}}

\begin{document}
\begin{titlepage}

\begin{center}
{\LARGE Matter Effects on Neutrino Oscillations in Long Baseline Experiments}\\
\vspace{1.1cm}
{\large Irina Mocioiu${}$\footnote{email: 
mocioiu@insti.physics.sunysb.edu }$^{(a)}$ 
and Robert Shrock${}$\footnote{email: 
shrock@insti.physics.sunysb.edu }$^{(a,b)}$}\\
\vspace{18pt}

(a) C.N. Yang Institute for Theoretical Physics \\
 State University of New York \\
 Stony Brook, NY 11794-3840 \\
(b) Physics Department \\
  Brookhaven National Laboratory \\
  Upton, NY  11973 \\
\end{center}
\vskip 0.6 cm

\begin{abstract}
\vspace{2cm}

We calculate matter effects on neutrino oscillations relevant for long baseline
experiments. In particular, we compare the results obtained with constant
density along the neutrino path versus results obtained by incorporating the
actual density profiles in the Earth.  We study the dependence of the
oscillation signal on both $E/\Delta m^2_{atm}$ and on the angles in the
leptonic mixing matrix. We also comment on the influence of $\Delta m^2_{sol}$
on the oscillations. The results show quantitatively how, as a function of
these input parameters, matter effects can cause significant ( 25 \%) changes
in the oscillation probabilities.  An important conclusion is that matter
effects can be useful in amplifying certain neutrino oscillation signals and
helping one to obtain measurements of mixing parameters and the magnitude and
sign of $\Delta m^2_{atm}$.
	


\end{abstract}
\end{titlepage}

\section{Introduction}

In a modern theoretical context, one generally expects nonzero neutrino masses
and associated lepton mixing.  Experimentally, there has been accumulating
evidence for such masses and mixing.  All solar neutrino experiments
(Homestake, Kamiokande, SuperKamiokande, SAGE, and GALLEX) show a significant
deficit in the neutrino fluxes coming from the Sun \cite{sol}. This deficit
can be explained by oscillations of the $\nu_e$'s into other weak
eigenstate(s), with $\Delta m^2_{sol}$ of the order $10^{-5}$ eV$^2$ for
solutions involving the Mikheev-Smirnov-Wolfenstein (MSW) resonant matter
oscillations \cite{wolf,ms} or of the order of $10^{-10}$ eV$^2$ for vacuum
oscillations.  Accounting for the data with vacuum oscillations (VO) requires
almost maximal mixing.  The MSW solutions include one for small mixing angle
(SMA) and one with essentially maximal mixing (LMA).

Another piece of evidence for neutrino oscillations is the atmospheric neutrino
anomaly, observed by Kamiokande \cite{kam}, IMB \cite{imb}, SuperKamiokande
\cite{sk} with the highest statistics, and by Soudan \cite{soudan} and MACRO
\cite{macro}.  This data can be fit by the inference of $\nu_{\mu} \rightarrow
\nu_x$ oscillations with $\Delta m^2_{atm}\sim 3.5 \times 10^{-3}$ eV$ ^2$
 and maximal mixing $\sin^2 2 \theta_{atm} = 1$ \cite{sk}.  The identification 
$\nu_x = \nu_\tau$ is preferred over $\nu_x=\nu_{sterile}$ at about the $2.5 
\sigma$ level \cite{learned}, and the identification $\nu_x=\nu_e$ is excluded 
by both the Superkamiokande data and the Chooz experiment \cite{chooz,exp}. 
  
In addition, the LSND experiment has reported observing $\bar\nu_\mu \to \bar
\nu_e$ and $\nu_{\mu} \to \nu_e$ oscillations with $\Delta m^2_{LSND} \sim 0.1
- 1$ eV$^2$ and a range of possible mixing angles, depending on $\Delta
m^2_{LSND}$ \cite{lsnd}. This result is not confirmed, but also not completely ruled out,
by a similar experiment, KARMEN \cite{karmen}.

There are currently intense efforts to confirm and extend the evidence for
neutrino oscillations in all of the various sectors -- solar, atmospheric, and
accelerator.  Some of these experiments are running; these include the Sudbury
Neutrino Observatory, SNO, and the K2K long baseline experiment between KEK and
Kamioka.  Others are in development and testing phases, such as BOONE, MINOS,
the CERN - Gran Sasso program, KAMLAND, and Borexino \cite{anl}.  Among the
long baseline neutrino oscillation experiments, the approximate distances are
$L \simeq 250$ km for K2K, 730 km for both MINOS, from Fermilab to Soudan and
the proposed CERN-Gran Sasso experiments.  The sensitivity of these experiments
is projected to reach down roughly to the level $\Delta m^2 \sim 10^{-3}
$eV$^2$.  There is strong motivation for another generation of experiments with
even higher sensitivity that can confirm the $\nu_\mu \to \nu_\tau$ transition
with the values of $\Delta m^2_{atm}$ and $\sin^2 2\theta_{atm}$ reported so
far and carry out further measurements of various neutrino oscillation
channels.

Recently, there has been considerable interest in the idea of a muon storage
ring that would serve as a ``neutrino factory'', i.e., a source of quite high
intensity, flavor-pure neutrino and antineutrino beams: $\nu_\mu + \bar\nu_e$
($\bar\nu_\mu + \nu_e$) from stored $\mu^-$'s ($\mu^+$'s) respectively 
\cite{geer}-\cite{bnlbook}.  Given the very high intensities anticipated to be
of order $10^{20}$ and perhaps even $10^{21}$ muon decays per year in various
preliminary studies, one can envision neutrino oscillation experiments with
quite long baselines of order several thousand km, with commensurate
sensitivity to various neutrino oscillation channels. 

One of the appeals of the muon storage ring/neutrino factory is that one can
measure several different neutrino oscillation transitions, using both the
$\nu_\mu$ ($\bar\nu_\mu$) and $\bar\nu_e$ ($\nu_e$) from a $\mu^-$ ($\mu^+$)
beam.  In this paragraph we assume a $\mu^-$ beam for definiteness (figures
below are shown for neutrinos from both stored $\mu^+$ and $\mu^-$ beams). In
addition to a high-statistics measurement of $\nu_\mu \to \nu_\mu$, as a
disappearance test for the $\nu_\mu \to \nu_\tau$ oscillation, one has also
various other channels.  Among these are $\bar\nu_e \to \bar\nu_\mu$, for which
the signal is a ``wrong-sign muon'', $\mu^+$, and $\bar\nu_e \to \bar\nu_\tau$,
which, in about 18 $\%$ of its decays, would also yield a wrong-sign muon.  The
measurement of the muon charge would be possible with either a magnetized iron
detector or a combination of a massive water \v{C}erenkov detector followed by
a muon spectrometer.  With sufficient detector capabilities, one could also
search for $\tau$ appearance, as is envisioned by the ICANOE and OPERA
experiments at Gran Sasso \cite{icanoe,opera}, although this requires neutrino
energies $E_\nu \gtwid 20$ GeV to avoid kinematic suppression of $\tau$
production.  

An important effect that must be taken into account in such experiments
concerns the matter-induced oscillations which neutrinos undergo along their
flight path through the Earth from the source to the detector.  Given the
typical density of the earth, matter effects are important for the neutrino
energy range $E \sim O(10)$ GeV and $\Delta m^2_{atm} \sim 3 \times 10^{-3}$
eV$^2$ values relevant for the long baseline experiments, in particular, for
the oscillation channels involving $\nu_e$, as we shall show below.  Matter
effects can also be important for the neutrino energy range $E \sim O(10)$ MeV
and $\Delta m^2 \sim 10^{-5}$ eV$^2$ involved in MSW solutions to the solar
neutrino problem.  After the initial discussion of matter-induced resonant
neutrino oscillations in \cite{wolf}, an early study of these effects including
three generations was carried out in \cite{barger80}.  The sensitivity of an
atmospheric neutrino experiment to small $\Delta m^2$ due to the long
baselines and the necessity of taking into account matter effects was discussed
e.g., in \cite{snowmass}.  After Ref. \cite{ms}, many analyses were performed
in the 1980's of the effects of resonant neutrino oscillations on the solar
neutrino flux, and matter effects in the Earth 
were studied, e.g., \cite{kp88} and
\cite{baltz}, which also discussed the effect on atmospheric neutrinos (see
also the review \cite{kimpev}).  Recent papers on matter effects relevant to
atmospheric neutrinos include \cite{petcov,akh}. Early studies of matter
effects on long baseline neutrino oscillation experiments were carried out in
\cite{bernpark}.  More recent analyses relevant to neutrino factories include
\cite{geer,dgh}, \cite{arubbia}-\cite{kim}.

 In this paper we shall present calculations of the
matter effect for parameters relevant to possible long baseline neutrino
experiments envisioned for the muon storage ring/neutrino factory.  In
particular, we compare the results obtained with constant density along the
neutrino path versus results obtained by incorporating the actual density
profiles.  We study the dependence of the oscillation signal on both $E/\Delta
m^2_{atm}$ and on the angles in the leptonic mixing matrix. We also comment on
the influence of $\Delta m^2_{sol}$ and CP violation on the oscillations. Some
of our results were presented in Ref. \cite{nnn99}.
Additional recent studies include \cite{cpv}-\cite{lindneretal}. 

In a hypothetical world in which there were only two neutrinos, $\nu_\mu$ and
$\nu_\tau$, the $\nu_\mu \to \nu_\tau$ oscillations in matter would be the same
as in vacuum, since both have the same forward scattering amplitude, via $Z$
exchange, with matter.  However, in the realistic case of three generations,
because of the indirect involvement of $\nu_e$ due to a nonzero $U_{13}$, and
because of the fact that $\nu_e$ has a different forward scattering amplitude
off of electrons, involving both $Z$ and $W$ exchange, there will be a
matter-induced oscillation effect on $\nu_\mu \to \nu_\tau$ (as well as other
channels).

We consider the usual three flavors of active neutrinos, with no light 
sterile (=electroweak-singlet) neutrinos.  This is sufficient to describe the
solar and atmospheric neutrino deficit.  If one were also to include the LSND
experiment, then, to obtain a reasonable fit, one would be led to include light
electroweak-singlet neutrinos.  Since the LSND experiment has not so
far been confirmed, we shall, while not prejudging the outcome of the BOONE
experiment, not include this in our fit.  We calculate
oscillation probabilities in the full $3\times 3$ mixing case and we study when
$\Delta m^2_{sol}$ can be relevant. In most cases there is only one mass scale
relevant for long baseline neutrino oscillations, $\Delta m^2_{atm} \sim {\rm
few} \times 10^{-3}$ eV$^2$ and we work with the hierarchy
\beq
\Delta m^2_{21}
= \Delta m^2_{sol} \ll \Delta m^2_{31} \approx \Delta m^2_{32}=\Delta m^2_{atm}
\label{hierarchy}
\eeq

In our work we take into account the actual profile of the Earth, as given by
geophysical seismic data \cite{prem} and compare the results with those
calculated using the approximation of average density along the path of the
neutrino.  Further, when only one mass squared difference is relevant, we
present the oscillation probabilities as functions of $E/\Delta m^2$, where
here and below, $E = E_\nu$.  This way of presenting the results is useful
since, for a given $L$ value, it shows the matter effect for a wide range of
$E$ and $\Delta m^2$ and hence can serve as an input in the choice of optimal
beam energy (along with other considerations such as the cross section
dependence $\sigma \sim E$ and the beam divergence $\sim (LE)^{-2}$, which,
together, favor higher values of $E$ to achieve a high event rate).  We study
how these oscillation probabilities vary with the different input parameters
and discuss the influence of the matter effects on the sensitivity to each of
these parameters.

\section{Theoretical Framework}

We first recall the form of the lepton mixing matrix.  
Let us denote the flavor vectors of SU(2) $\times$ U(1)
nonsinglet neutrinos as $\nu = (\nu_e,\nu_\mu,\nu_\tau)$ and the 
vector of electroweak-singlet neutrinos as $\chi = (\chi_1,..,\chi_{n_s})$.  
The Dirac and Majorana neutrino mass terms can then be written compactly as 
\beq
-{\cal L}_m =
 {1 \over 2}(\bar\nu_L \ \overline{\chi^c}_L)
             \left( \begin{array}{cc}
              M_L & M_D \\
              (M_D)^T & M_R \end{array} \right )\left( \begin{array}{c}
      \nu^{c}_R \\
      \chi_R \end{array} \right ) + h.c.
\label{numass}
\eeq
where $M_L$ is the $3 \times 3$ left-handed Majorana mass matrix, $M_R$ is a
$n_s \times n_s$ right-handed Majorana mass matrix, and $M_D$ is the 3-row by
$n_s$-column Dirac mass matrix.  In general, all of these are complex, and
$(M_L)^T = M_L \ , \quad (M_R)^T = M_R$.  Without further theoretical input,
the number $n_s$ of electroweak singlet neutrinos is not determined.  For
example, in the minimal SU(5) grand unified theory (GUT), $n_s=0$, while in
SO(10), $n_s=3$. Within this theoretical context, since the terms $\chi_{j
R}^TC \chi_{k R}$ are electroweak singlets, the associated coefficients, which
comprise the elements of the matrix $M_R$, would not be expected to be related
to the electroweak symmetry breaking scale, but instead, would be expected to
be much larger, plausibly of order the GUT scale.  Furthermore, the
left-handed Majorana mass terms can only arise via operators of dimension at
least 5, such as
\beq
{\cal O} = \frac{1}{M_X}\sum_{a,b}h_{a,b}
(\epsilon_{ik}\epsilon_{jm}+\epsilon_{im}\epsilon_{jk})
\Bigl [ {\cal L}^{Ti}_{a L}C {\cal L}^j_{b L} \Bigr ] \phi^k \phi^m + h.c.
\label{majleft}
\eeq
where ${\cal L}_{L a} = (\nu_{\ell_a}, \ell_a)_L^T$ is the left-handed,
$I=1/2$, $Y=-1$ lepton doublet with
generation index $a$ ($a=1$, 2, or 3), where 
$\ell_{a}=e, \mu, \tau$, for $a=1,2,3$,
$M_X$ denotes a generic mass scale characterizing the origin of this term, and
$\phi$ is the standard model Higgs or $H_u$ in the supersymmetric standard
model.  Because (\ref{majleft}) is a nonrenormalizable operator, the success of
the standard model as a renormalizable field theory then implies that $M_X$ is
much larger than the scale of electroweak symmetry breaking, and, within a GUT
context, $M_X$ would be of order the GUT scale, as with $M_R$. 
The terms arising from the vacuum expectation values
of the Higgs doublets then make up the submatrix $M_L$.  The resultant
diagonalization of the matrix in eq. (\ref{numass}) then naturally leads to a
set of 3 light masses for the three known neutrinos, generically of order
$m_\nu \sim m_D^2/M_R$, and $n_s$ very large masses generically of order $M_R$,
for the electroweak singlet neutrinos.  This seesaw mechanism is very
appealing, since it can provide a plausible explanation for why the known
neutrinos are so light \cite{seesaw}.  Although the full leptonic mixing matrix
is $(3+n_s) \times (3+n_s)$ dimensional, the light and heavy neutrinos largely
decouple from each other so that, to a high degree of accuracy, one can
describe the linear combinations of the $(3+n_s)$ mass eigenstates that form
$(3+n_s)$ weak eigenstates in a decoupled manner, using a simple $3 \times 3$
matrix $U$, which, to high accuracy, is unitary, for the known neutrinos.  This
is determined by the diagonalization of the effective $3 \times 3$ light
neutrino mass matrix
\beq
M_\nu = - M_D M_R^{-1} M_D^T
\label{meffective}
\eeq
and an $n_s \times n_s$ matrix for the heavy neutrino sector, which matrix will
not be used here.  The lepton mixing matrix can then be written as the unitary
matrix 
\beqs
U=R_{23}KR_{13}K^*R_{12}K'=
\pmatrix{c_{12} c_{13} & s_{12}c_{13} & s_{13} e^{-i\delta} \cr 
-s_{12}c_{23}-c_{12}s_{23}s_{13}e^{i\delta}
& c_{12}c_{23}-s_{12}s_{23}s_{13}e^{i\delta} & s_{23}c_{13} \cr 
s_{12}s_{23}-c_{12}c_{23}s_{13}e^{i\delta}
&-c_{12}s_{23}-s_{12}c_{23}s_{13}e^{i\delta} & c_{23}c_{13}}K'
\eeqs
where $R_{ij}$ is the rotation matrix in the $ij$ subspace, 
$c_{ij}=\cos\theta_{ij}$, $s_{ij}=\sin\theta_{ij}$, 
$K=diag(e^{-i\delta},1,1)$ and $K'$ involves further possible phases due to
Majorana mass terms that will contribute here.  

In passing, we note that although this theoretical context is appealing,
various modifications are possible.  For example, string theory generically
involves certain moduli fields which are singlets under the standard model
gauge group, have flat superpotentials, and hence are massless in perturbation
theory down to the energy scale where supersymmetry is broken.  The spinor
component fields, modulinos, can act as electroweak-singlet neutrinos, and may
well have masses much less than the GUT scale (e.g. \cite{benakli}).  Moreover,
models with a low string scale $<< M_{Planck}$ and large compact dimensions
(e.g., \cite{dd}) also have implications for neutrino phenomenology.  Here we
shall work within the conventional seesaw-type scenario because of its
simplicity and success in accounting for the most striking known feature of
neutrinos, namely the fact that they are so light compared with the other known
fermions.

For our later discussion it will be useful to record the formulas for the
various relevant neutrino oscillation transitions.  In the absence of any
matter effect, the probability that a (relativistic) weak neutrino eigenstate
$\nu_a$ becomes $\nu_b$ after propagating a distance $L$ is 
\beqs
P(\nu_a \to \nu_b) &=& \delta_{ab} - 4 \sum_{i>j=1}^3 
Re(K_{ab,ij}) \sin^2  \Bigl ( \frac{\Delta m_{ij}^2 L}{4E} \Bigr ) + 
\nonumber\\&+& 4 \sum_{i>j=1}^3 Im(K_{ab,ij})
 \sin \Bigl ( \frac{\Delta m_{ij}^2 L}{4E} \Bigr ) 
\cos \Bigl ( \frac{\Delta m_{ij}^2 L}{4E} \Bigr ) 
\label{pab}
\eeqs
where
\beq
K_{ab,ij} = U_{ai}U^*_{bi}U^*_{aj} U_{bj}
\label{k}
\eeq
and
\beq
\Delta m_{ij}^2 = m(\nu_i)^2-m(\nu_j)^2
\label{delta}
\eeq
Recall that in vacuum, CPT invariance implies 
$P(\bar\nu_b \to \bar\nu_a)=P(\nu_a \to \nu_b)$ and hence, for $b=a$, 
$P(\bar\nu_a \to \bar\nu_a) = P(\nu_a \to \nu_a)$.  For the 
CP-transformed reaction $\bar\nu_a \to \bar\nu_b$ and the T-reversed
reaction $\nu_b \to \nu_a$, the transition probabilities are given by the
right-hand side of (\ref{pab}) with the sign of the imaginary term reversed. 
(Below we shall assume CPT invariance, so that CP violation is equivalent to T
violation.) For most sets of parameters, only one mass scale is relevant for 
the neutrino oscillations of interest here, namely 
\beq
\Delta m^2_{atm} = \Delta m^2_{32} 
\label{deltaatm}
\eeq
In this case, CP (T) violation effects are negligibly small, so that in 
vacuum 
\beq
P(\bar\nu_a \to \bar\nu_b) = P(\nu_a \to \nu_b) 
\label{pcp}
\eeq
\beq
P(\nu_b \to \nu_a) = P(\nu_a \to \nu_b)
\label{pt}
\eeq
In the absence of T violation, the second equality (\ref{pt}) would still hold
in matter, but even in the absence of CP violation, the first equality
(\ref{pcp}) would not hold.  With the hierarchy (\ref{hierarchy}), the
expressions for the specific oscillation transitions are 
\beqs
P(\nu_\mu \to \nu_\tau) & = & 4|U_{33}|^2|U_{23}|^2
\sin^2 \Bigl ( \frac{\Delta m^2_{atm}L}{4E} \Bigr ) \cr\cr
& = & \sin^2(2\theta_{23})\cos^4(\theta_{13})
\sin^2 \Bigl (\frac{\Delta m^2_{atm}L}{4E} \Bigr )
\label{pnumunutau}
\eeqs

\beqs
P(\nu_e \to \nu_\mu) & = & 4|U_{13}|^2 |U_{23}|^2 
\sin^2 \Bigl ( \frac{\Delta m^2_{atm}L}{4E} \Bigr ) \cr\cr
& = & \sin^2(2\theta_{13})\sin^2(\theta_{23}) 
\sin^2 \Bigl (\frac{\Delta m^2_{atm}L}{4E} \Bigr )
\label{pnuenumu}
\eeqs

\beqs
P(\nu_e \to \nu_\tau) & = & 4|U_{33}|^2 |U_{13}|^2
\sin^2 \Bigl ( \frac{\Delta m^2_{atm}L}{4E} \Bigr ) \cr\cr
& = & \sin^2(2\theta_{13})\cos^2(\theta_{23}) 
\sin^2 \Bigl (\frac{\Delta m^2_{atm}L}{4E} \Bigr )
\label{pnuenutau}
\eeqs

In neutrino oscillation searches using reactor antineutrinos,
i.e. tests of $\bar\nu_e \to \bar\nu_e$, the two-species mixing hypothesis used
to fit the data is 
\beq
P(\nu_e \to \nu_e) = 1 - \sin^2(2\theta_{reactor})
\sin^2 \Bigl (\frac{\Delta m^2_{reactor}L}{4E} \Bigr )
\label{preactor}
\eeq
where $\Delta m^2_{reactor}$ is the squared mass difference relevant for 
$\bar\nu_e \to \bar\nu_x$.  In particular, in the upper range of values of
$\Delta m^2_{atm}$, since the transitions $\bar\nu_e \to \bar\nu_\mu$ and
$\bar\nu_e \to \bar\nu_\tau$ contribute to $\bar\nu_e$ disappearance, one has 
\beq
P(\nu_e \to \nu_e) = 1 - \sin^2(2\theta_{13})\sin^2 \Bigl 
(\frac{\Delta m^2_{atm}L}{4E} \Bigr )
\label{preactoratm}
\eeq
i.e., $\theta_{reactor}=\theta_{13}$, and the Chooz reactor experiment yields 
the bound \cite{chooz} 
\beq
\sin^2(2\theta_{13}) < 0.10
\label{chooz}
\eeq
which is also consistent with conclusions from the SuperK data analysis 
\cite{sk}. 

Further, the quantity ``$\sin^2(2\theta_{atm})$'' often used to fit
the data on atmospheric neutrinos with a simplified two-species mixing
hypothesis, is, in the three-generation case, 
\beq
\sin^2(2\theta_{atm}) \equiv \sin^2(2\theta_{23})\cos^4(\theta_{13})
\label{thetaatm}
\eeq
Hence for small $\theta_{13}$, as implied by (\ref{chooz}), it follows that, to
good accuracy, $\theta_{atm} = \theta_{23}$.

\section{Calculation of Matter Effects} 

The evolution of the flavor eigenstates of neutrinos is given by
\beq
{\sl i}\frac{\sl d}{\sl d x}\nu =\left(\frac{1}{2E}UM^2U^{\dagger}+V\right)\nu
\eeq
where
\beq
\nu=U\nu_m
\eeq
\beq
\nu_m=\pmatrix{\nu_1\cr \nu_2\cr\nu_3}
\eeq
\beq
M^2=\pmatrix{m_1^2&0&0\cr0&m_2^2&0\cr0&0&m_3^2} \quad , 
V=\pmatrix{\sqrt{2}G_FN_e&0&0\cr0&0&0&\cr0&0&0}
\label{V}
\eeq
Here $N_e$ is the electron number density and we have 
$\sqrt{2}G_FN_e$ [eV]$=7.6\times 10^{-14} Y_e \rho$ [g/cm$^3$].

The atmospheric neutrino data suggests almost maximal mixing in the $2-3$
sector. However, a small but non-zero $s_{13}$ is still allowed, and this
produces the matter effect in the traversal of neutrinos through the Earth.  
We use the bound (\ref{chooz}) on $\sin^2(2\theta_{13})$ here, consistent with
both the Chooz experiment \cite{chooz} and the atmospheric neutrino data 
\cite{sk}.

If we assume that the solar neutrino deficiency is explained by the small
mixing angle (SMA) MSW solution or by vacuum oscillations, with the hierarchy
of eq.  (\ref{hierarchy}), it follows that, for the relevant energies $E \gtwid
1$ GeV and path-lengths $L \sim 10^3 - 10^4$ km, only one squared mass scale,
$\Delta m^2_{atm}$, is important for the oscillations and the three-species
neutrino oscillations can be described in terms of this quantity, $\Delta
m^2_{atm}$, and the mixing parameters $\sin^2(2\theta_{23})$, and
$\sin^2(2\theta_{13})$, with negligible dependence on $\sin^2(2\theta_{12})$
and $\delta$.

In order to write down the probabilities of oscillation for long-baseline 
and atmospheric neutrinos, it is convenient to transform to a new basis 
defined by (e.g. \cite{akh}) 

\beq
\nu=R_{23}\tilde\nu
\eeq
The evolution of $\tilde\nu$ is given by 
\beq
\tilde H = \frac{1}{2E} K R_{13}K^* R_{12}M^2 R_{12}^{\dagger} K 
R_{13}^{\dagger} K^* +V
\eeq
In the one mass-scale approximation, this can be reduced to
\beq
\tilde H \simeq \pmatrix{\frac{1}{2E}s_{13}^2\Delta m_{32}^2+\sqrt{2}G_FN_e&0&
\frac{1}{2E}s_{13}c_{13}\Delta m_{32}^2 e^{-i\delta}\cr
0&\frac{1}{2E}c_{12}^2\Delta m_{21}^2&0\cr
\frac{1}{2E}s_{13}c_{13}\Delta m_{32}^2 e^{i\delta}
&0&\frac{1}{2E}c_{13}^2\Delta m_{32}^2}
\eeq
It can be seen now that in the basis $(\nu_e,\tilde\nu_\mu,\tilde\nu_\tau)$ the
three-flavor evolution equation decouples and it is enough to treat the
two-flavor case.  We define $S$ and $P$ by 
\beq
\pmatrix{\nu_e\cr \tilde\nu_\mu\cr\tilde\nu_\tau}(x)=S
\pmatrix{\nu_e\cr\tilde\nu_\mu\cr\tilde\nu_\tau}(0)
\eeq
and
\beq
P\equiv |S_{13}|^2=1-|S_{33}|^2
\eeq
Transforming back to the flavor 
basis $(\nu_e,\nu_\mu,\nu_\tau)$, the probabilities of oscillation become 

\beqs
P(\nu_e\rightarrow\nu_e)&=&1-P\\
P(\nu_e\rightarrow\nu_\mu)&=& P(\nu_\mu\rightarrow\nu_e)=s_{23}^2 P\\
P(\nu_e\rightarrow\nu_\tau)&=&c_{23}^2 P\\
P(\nu_\mu\rightarrow\nu_\mu)&=&1-s_{23}^4P + 2 s_{23}^2c_{23}^2 
[Re(S_{22}S_{33})-1]\\
P(\nu_\mu\rightarrow\nu_\tau)&=&s_{23}^2c_{23}^2[2-P-2Re(S_{22}S_{33})]
\eeqs

Note that for the mass hierarchy (\ref{hierarchy}), the CP-violating phases
disappear from the oscillation probabilities.  In this case what we need to
solve is the evolution equation for a two-flavor neutrino system.  By
subtracting from the diagonal the quantity 
$\frac{1}{4E}\Delta m^2_{32} + \frac{1}{\sqrt{2}} G_FN_e$, this can be 
written in the form
\beq
{\sl i}\frac{\sl d}{\sl d x}\pmatrix{\nu_a \cr \nu_b} = \pmatrix {-A(x) & B 
\cr B&A(x)}
\pmatrix{\nu_a\cr\nu_b}
\label{evol}
\eeq
with
\beq
A(x)=\frac{\Delta m^2_{32}}{4E}\cos(2\theta_{13})-\frac{G_F}{\sqrt2}N_e(x)
\eeq
\beq
B=\frac{\Delta m^2_{32}}{4E}\sin(2\theta_{13})
\eeq

\begin{figure}                             
\begin{center}
\mbox{\epsfxsize=10truecm
\epsffile{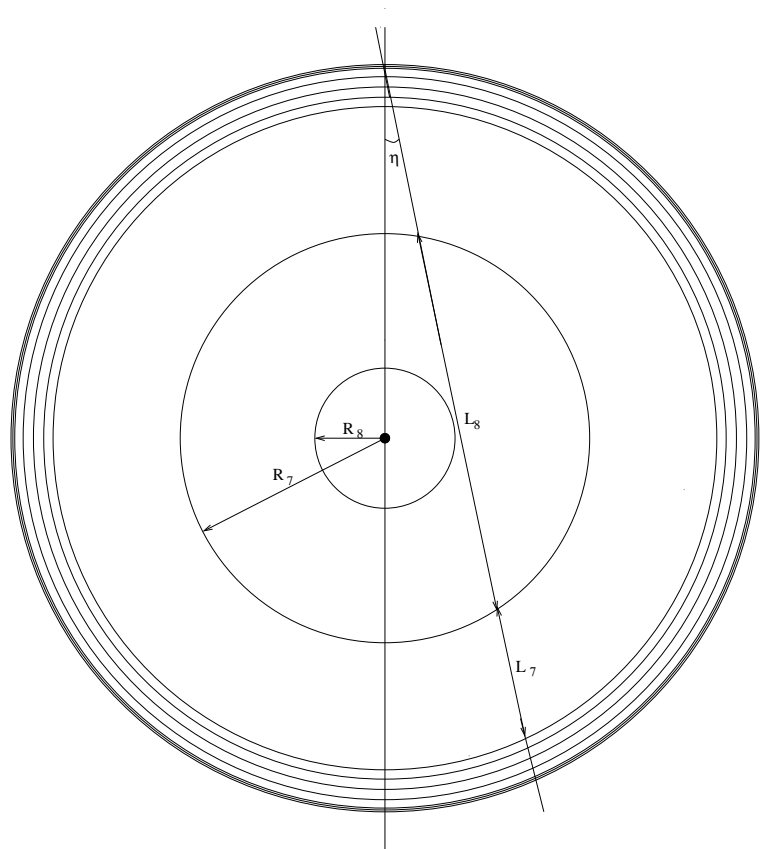}}
\mbox{\epsfxsize=10truecm
\epsffile{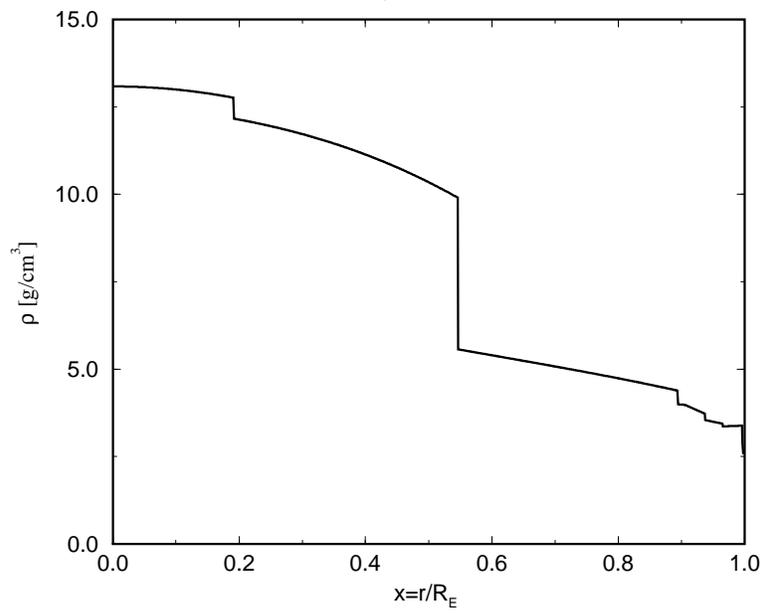}}
\end{center}
 \caption{\footnotesize{Density profile of the Earth.}}
\label{fig:density}
\end{figure}

For our purposes, we recall that the Earth is composed of crust, mantle, liquid
outer core, and solid inner core, together with additional sublayers in the
mantle, with particularly strong changes in density between the lower mantle
and outer core. The density profile of the Earth is shown in 
Fig.\ref{fig:density}. The densities of the different layers are given in 
Table 1 as 
function of the normalized radius $x={R}/{R_E}$, $R_E=6371$ km being the 
radius of the Earth.
 The core has average density
$\rho_{core}=11.83$ g/cm$^3$ and electron fraction $Y_{e,core}=0.466$, while
the mantle has average density $\rho_{mantle}=4.66$ g/cm$^3$ and
$Y_{e,mantle}=0.494$.  

\begin{center}
\begin{tabular}{|c|c|}
\hline
Radius [Km]&Density [g/cm$^3$]\\
\hline
$0-1221.5$&$13.0885-8.8381 x^2$\\
\hline
$1221.5-3480.0$&$12.5815-1.2638 x-3.6426 x^2-5.5281 x^3$\\
\hline
$3480.0-5701.0$&$7.9565-6.4761 x+5.5283 x^2 -3.0807 x^3$\\
\hline
$5701.0-5771.0$&$5.3197-1.4836 x$\\
\hline
$5771.0-5971.0$&$11.2494-8.0298 x$\\
\hline
$5971.0-6151.0$&$7.1089-3.8045 x$\\
\hline               
$6151.0-6346.6$&$2.6910+0.6924 x$\\
\hline
$6346.6-6356.0$&$2.900$\\
\hline
$6356.0-6371.0$&$2.600$\\
\hline
\end{tabular}
\end{center}
\begin{center}
{\bf Table 1} Density Profile of the Earth 
\end{center}

Since, to very good accuracy, the Earth is spherically symmetric 
(Fig.\ref{fig:density}), the neutrino flight path is 
described only by the zenith angle $\theta_z$ (or $\eta=\pi-\theta_z$). 
For
\be
\frac{R_{i+1}}{R}<\sin\eta<\frac{R_{i}}{R}
\ee 
the neutrinos pass 
through $2i+1$ layers in the Earth. The distances traveled by the neutrinos 
in each of these layers are

\ba
&\!& L_1 = R\cos\eta -\sqrt{R_1^2-R^2\sin^2\eta}\\
&\!& L_k = L_{2i+2-k}=\sqrt{R_{k-1}^2-R^2\sin^2\eta}-
\sqrt{R_k^2-R^2\sin^2\eta}\,\, ,
\,\,\,\,\,\,\,\, 2\leq k\leq i\\
&\!& L_{i+1}=2\sqrt{R_i^2-R^2\sin^2\eta}
\ea

Studies have been done using the  average density of the Earth along the 
neutrino path. In this case the evolution equation can be easily solved and 
the probability of oscillation is given by
\beq
P(\nu_a\rightarrow\nu_b)= \sin^2(2\theta_m) \sin^2(\omega L)
\eeq
where $\omega=\sqrt{A^2+B^2}$ and $\theta_m$ is the effective mixing in matter
given by
\beq
\sin^2(2\theta_m)=\frac{\sin^2(2\theta)}{\sin^2(2\theta)+
\Bigl ( \cos(2\theta) - \frac{2\sqrt{2}G_FN_e E}{\Delta m^2} \Bigr )^2 }
\label{thetam}
\eeq
Just as was the case with the application of the MSW analysis to solar
neutrinos, the key observation is that although the angle $\theta$, which is 
essentially $\theta_{13}$ here, is small, the vanishing of the term in
parentheses in the denominator of (\ref{thetam}) renders the effective mixing
angle $\theta_m=\pi/4$, thereby producing maximal mixing in matter.  The
important point is that, given the range of densities in the layers of the
Earth, and the value of $\Delta m_{atm}^2 \simeq 3.5 \times 10^{-3}$ eV$^2$,
this matter resonance occurs for neutrino energies of order $O(10)$ GeV, in 
the range planned for long baseline neutrino oscillation experiments. 
Since this effect clearly depends on the sign of $\Delta m^2$, the measurement
of matter effects can give information on this sign. 
When one takes account of the actual variable-density situation in the Earth,
it is necessary to perform a numerical integration of the evolution equation,
which we have done.  We also go beyond the one mass-scale approximation and
study the effect of $\Delta m^2_{sol}$ and $\theta_{12}$ on the oscillations.
In this case we calculate the oscillation probabilities for nonzero values of 
all six oscillation parameters (three angles, one phase, and two mass square 
differences) and discuss when the simpler cases are very good approximations.

\section{Results and Discussion}

\begin{figure}[hbtp]
\begin{center}
\mbox{\epsfxsize=8truecm
\epsfysize=8.6truecm
\epsffile{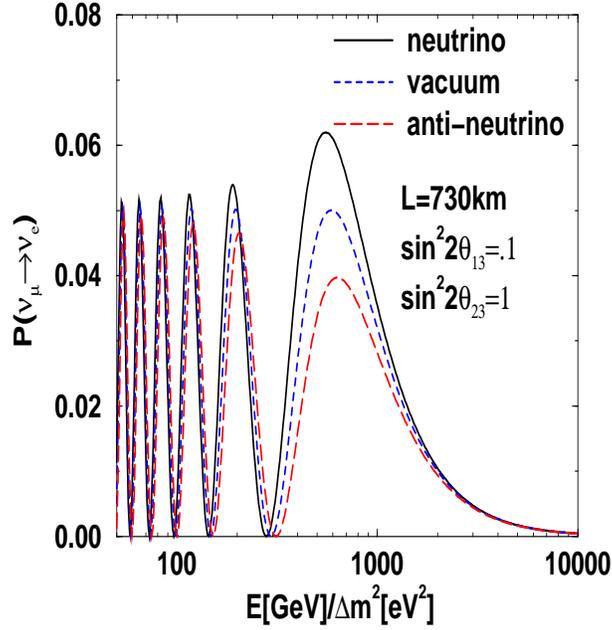}}
\end{center}
\caption{\footnotesize{$P(\nu_\mu\rightarrow \nu_e)$ and
$P(\bar\nu_\mu\rightarrow\bar\nu_e)$ in matter and
in vacuum for $L=730$ km.  (Figure legend is understood to include both cases.)
Here $\sin^2(2\theta_{13})=0.1$ and $\sin^2(2\theta_{23})=1$.}}
\label{fig:minos}
\end{figure}

  For long baseline experiments like K2K, MINOS, and CERN to
Gran-Sasso, the neutrino flight path only goes through the upper mantle.  The
density in this region is practically constant, and the oscillation
probabilities can easily be calculated. The matter effects are small, but
possibly detectable for the longer baselines.  We show in Fig.\ref{fig:minos}
an example for $P(\nu_\mu\rightarrow \nu_e)$ relevant for MINOS or the CERN to
Gran-Sasso experiments (if SMA or VO solve the solar problem).

However, there are several motivations for very long baseline experiments,
since, with sufficiently high-intensity sources, these can be sensitive to
quite small values of $\Delta m^2$ and since the matter effects, being larger,
can amplify certain oscillations and can, in principle, be used to get
information on the sign of $\Delta m^2_{atm}$.  Hence we concentrate here on
these very long baseline experiments; for these, the neutrino flight path goes
through several layers of the Earth with different densities, including the
lower mantle for some.  We show results for the Fermilab to SLAC path length
$L\simeq 2900$ km and for $L\simeq 7330$ km, the distance from Fermilab to Gran
Sasso. Path lengths corresponding to the distance BNL to SLAC, $L\simeq 4500$
km, and BNL to Gran Sasso, $L\simeq 6560$ km, are also considered.  We have
also performed calculations for $L \simeq$ 9200 km, the Fermilab to
SuperKamiokande path length.

\begin{figure}[hbtp]
\begin{center}
\mbox{\epsfxsize=8truecm
\epsfysize=8.4truecm
\epsffile{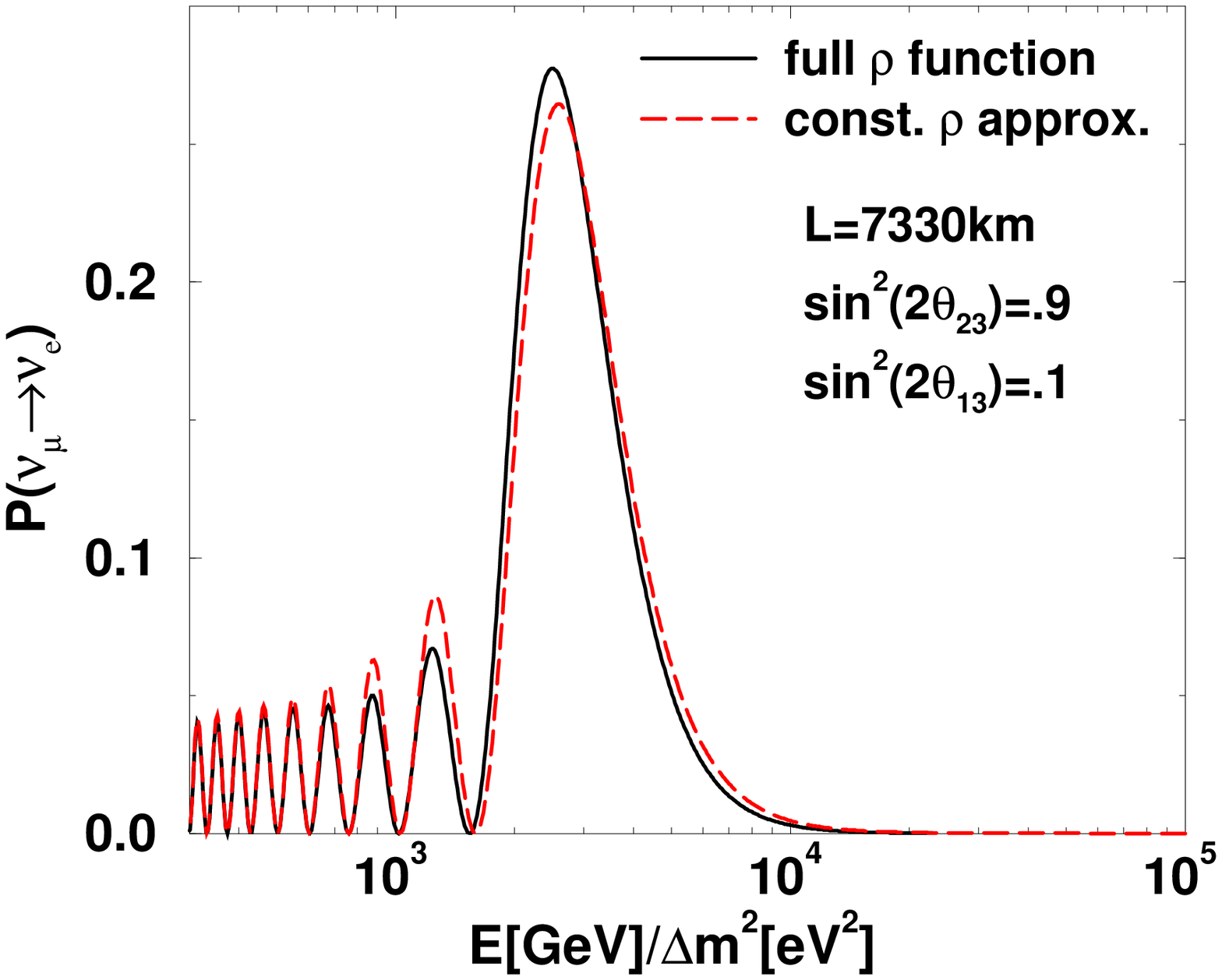}}
\mbox{\epsfxsize=8truecm
\epsfysize=8.6truecm
\epsffile{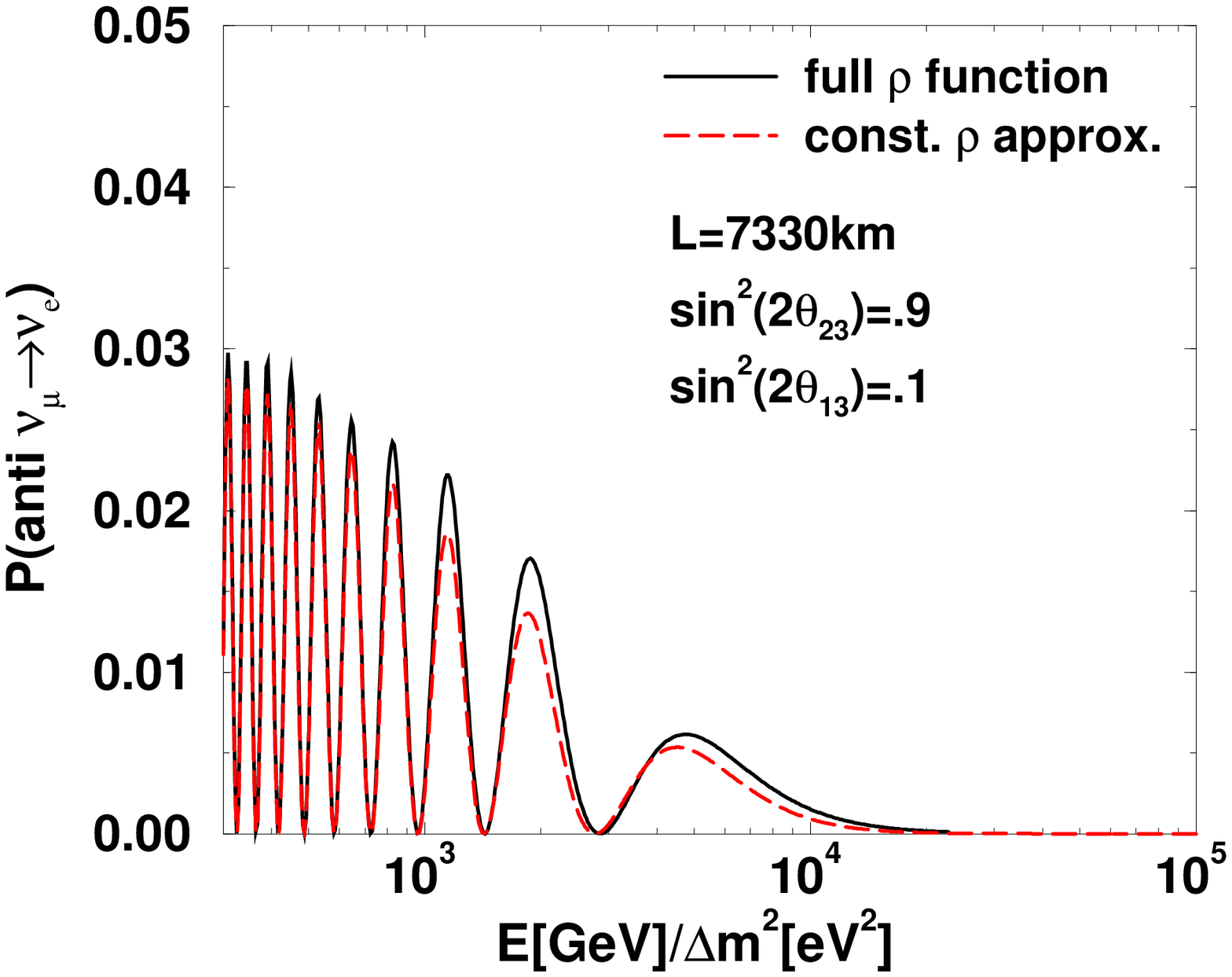}}
\end{center}
 \caption{\footnotesize{$P(\nu_\mu\rightarrow \nu_e)$ and
$P(\bar\nu_\mu\rightarrow\bar\nu_e)$ with density function given by
the full model of the Earth, compared with constant density approximation, 
for $L=7330$ km with $\sin^2(2\theta_{13})=0.1$ and 
$\sin^2(2\theta_{23})=0.9$.}}
\label{fig:mue19}
\end{figure}

        In Fig.\ref{fig:mue19} we compare the probabilities calculated with
constant density along the neutrino path versus the results obtained by
numerically integrating the evolution equation with the actual density profile
of the Earth, as given by \cite{prem}. The results are almost the same for most
of the parameter range. However, at given energies, as for example for the
second maximum in $P(\nu_\mu\rightarrow\nu_e)$, the correction to the
probability is of the order of 20\%. The results in Fig.\ref{fig:mue19} are
obtained for the $L=7330$ km distance, for which the beam goes through all
layers of the mantle.  In \cite{nnn99} we gave a series of similar comparisons
of oscillation probabilities calculated with the constant density approximation
and with the actual density function.  In the following, we only
present results obtained with the actual density function in the Earth.

\begin{figure}[hbtp]
\begin{center}
\mbox{\epsfxsize=8truecm
\epsfysize=8.6truecm
\epsffile{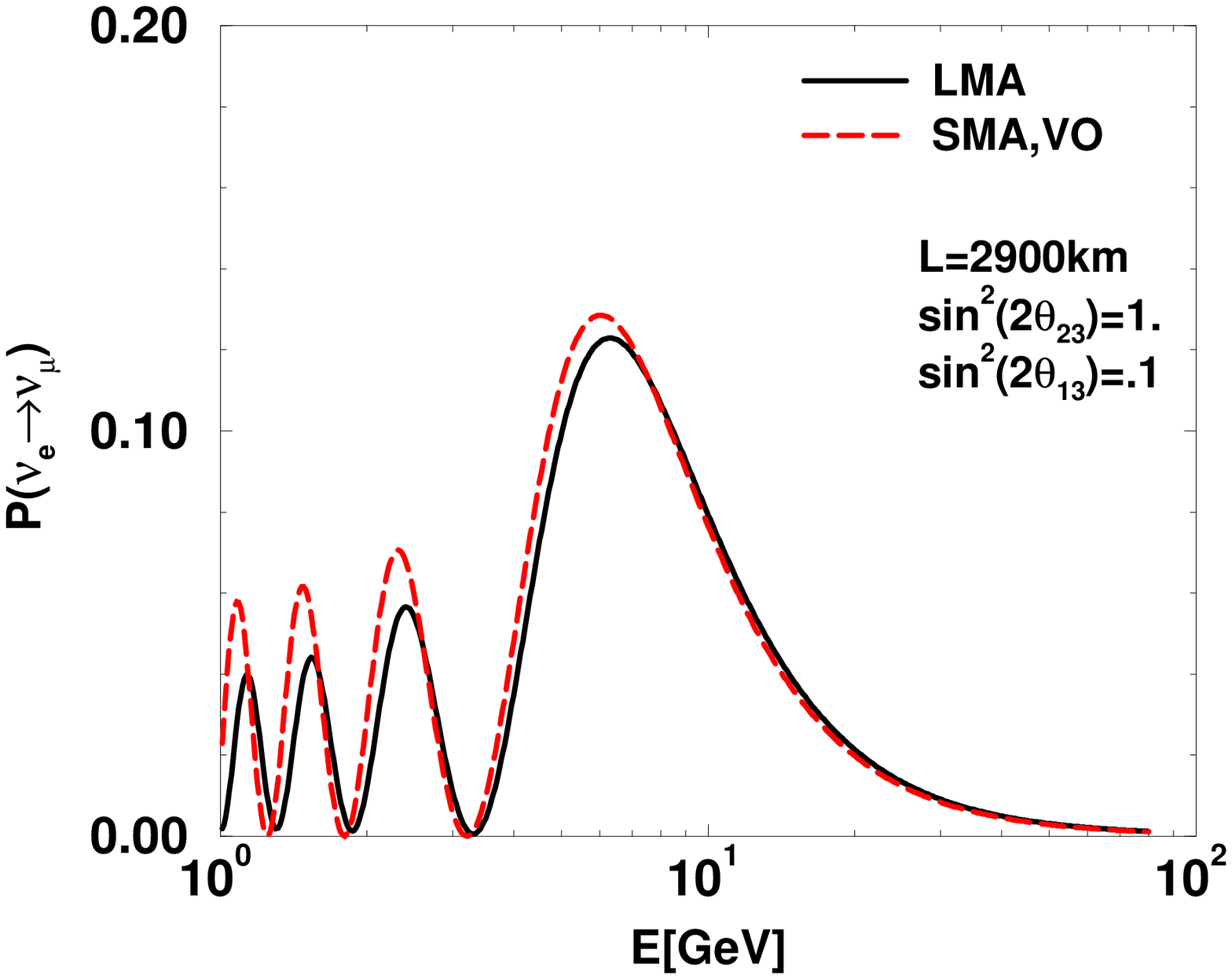}}
\mbox{\epsfxsize=8truecm
\epsfysize=8.6truecm
\epsffile{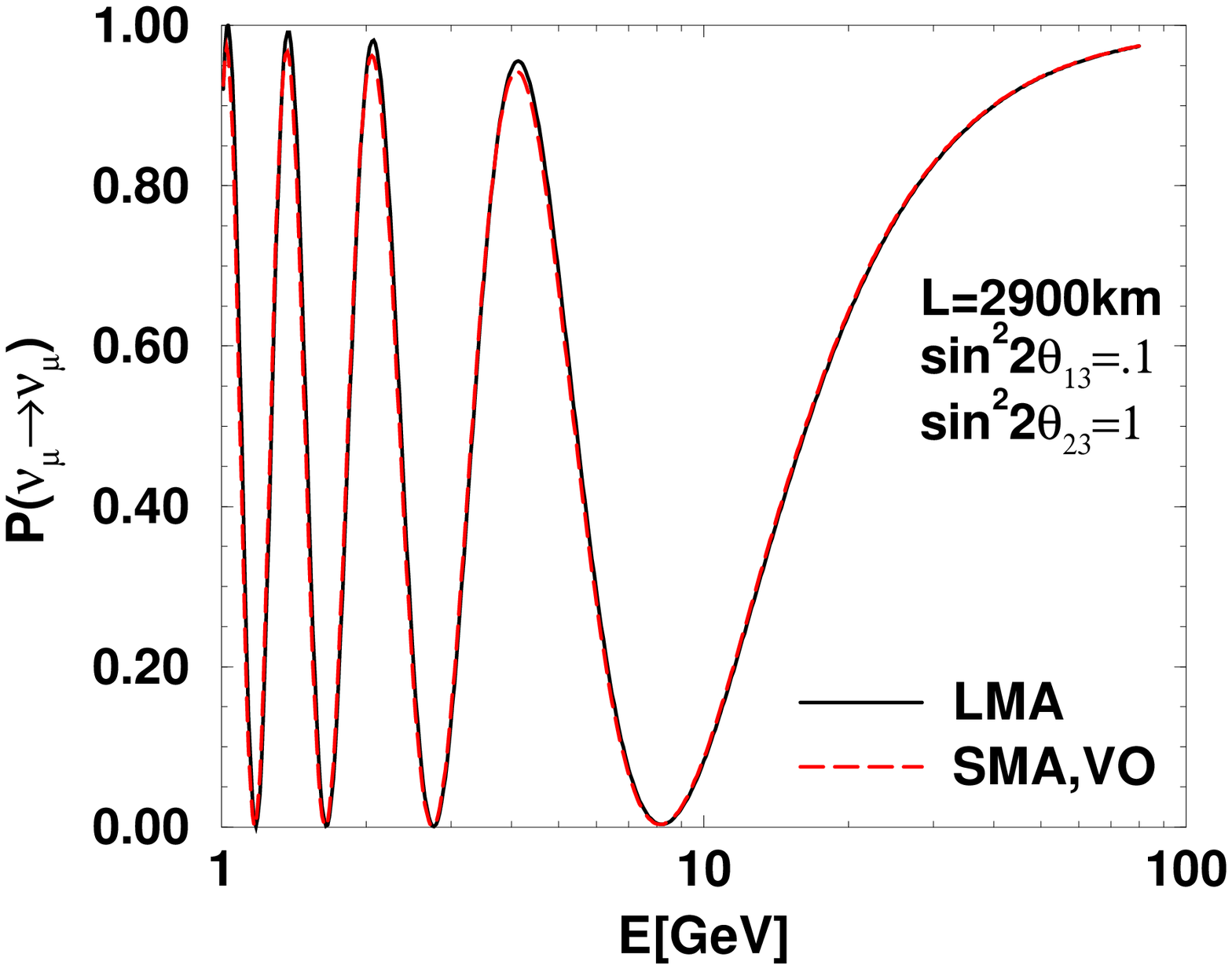}}
\end{center}
\caption{\footnotesize{$P(\nu_\mu\rightarrow \nu_e)$ and
$P(\nu_\mu \to \nu_\mu)$ for various solutions to the solar
neutrino problem. Here $L=2900$ km, $\sin^2(2\theta_{13})=0.1$,
$\sin^2(2\theta_{23})=1$, and $\Delta m^2_{atm}=3.5\times 10^{-3}$eV$^2$.}}
\label{fig:slac}
\end{figure}

If one assumes the LMA solution to the solar neutrino problem, then for the
$L=2900$ km baseline, both $\Delta m^2_{atm}$ and $\Delta m^2_{sol}$ have to be
considered.  In Fig.\ref{fig:slac} we show $P(\nu_\mu\rightarrow \nu_e)$ and
the $\nu_\mu$ survival probability $P(\nu_\mu \to \nu_\mu)$ as functions of
energy for $\Delta m^2_{atm}=3.5 \times 10^{-3}$ eV$^2$ and $\sin^2
2\theta_{23}=1$, as suggested by the atmospheric data, and
$\sin^22\theta_{13}=0.1$, the maximum value allowed, and two different choices
of $\Delta m^2_{sol}$ and $\sin^22\theta_{12}$. One choice corresponds to the
LMA solution, with $\Delta m^2_{sol}=5\times 10^{-5}$ eV$^2$ and
$\sin^22\theta_{12}=0.8$. For this LMA case, the choice of the CP violating
phase $\delta$ is relevant; here we take $\delta=0$ and compare with nonzero
$\delta$ below.  The other choice is for the VO solution, with $\Delta
m^2_{sol}=10^{-10}$ eV$^2$ and $\sin^22\theta_{12}=1$.  The SMA solution gives
the same results as the VO solution.  One sees that the terms
involving $\Delta m^2_{sol}$ can have non-negligible effects on the
$\nu_\mu\rightarrow\nu_e$ oscillation probability for this path-length,
especially at lower energies.

\begin{figure}[hbtp]
\begin{center}
\mbox{\epsfxsize=8truecm
\epsfysize=8.6truecm
\epsffile{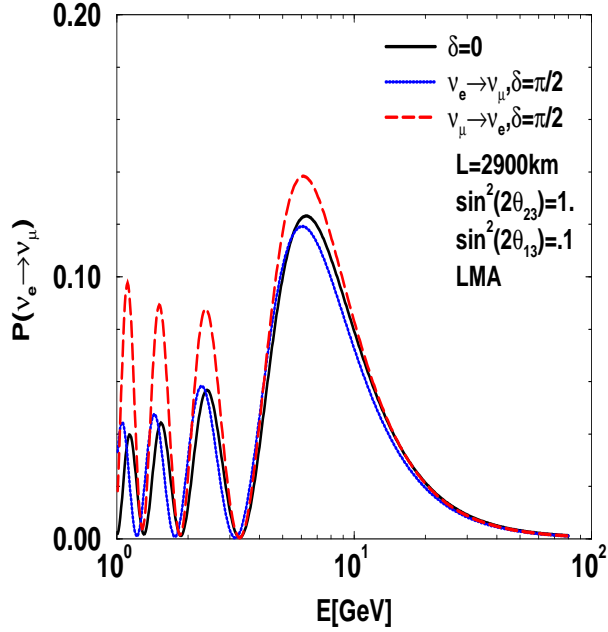}}
\end{center}
\caption{\footnotesize{CP-violation effects for $L=2900$ km, 
$\sin^2(2\theta_{23})=1$, $\sin^2(2\theta_{13})=0.1$, LMA solution, 
$\delta=0,\pi/2$.}}
\label{fig:cp}
\end{figure}

As noted earlier, in the one-mass scale approximation, there are no CP
violation effects in these oscillations; however, when we take into account
$\Delta m^2_{sol}$, we also have to consider CP-violating effects.  We present
in Fig.\ref{fig:cp} a comparison showing the results for the probabilities
$P(\nu_\mu\rightarrow\nu_e)$ and $P(\nu_e\rightarrow\nu_\mu)$ for $\delta=0$
and $\delta=\pi/2$. We consider $L=2900$ km, $\sin^22\theta_{23}=1$,
$\sin^22\theta_{13}=0.1$ and the LMA solution for the solar neutrino
problem. We can see that the effects of the CP-violating phase are small. Note
however that for non-zero CP-violation, $P(\nu_\mu\rightarrow \nu_e)\ne
P(\nu_e\rightarrow\nu_\mu)$.  For no CP-violation, even with matter effects,
there is no difference between these two probabilities.  Since with a muon
storage ring, by switching between $\mu^-$ and $\mu^+$ beams, one could obtain
both $\nu_\mu$ and $\nu_e$ beams, there is the possibility of searching for 
the CP (actually T) violating difference
$P(\nu_\mu\rightarrow\nu_e)- P(\nu_e \rightarrow\nu_\mu)$. 
In practice, however,
it would be difficult to identify the $e^-$ from $\nu_e$, given that the
$\mu^-$ stored beam that yields the initial $\nu_\mu$ also yields $\bar\nu_e$,
which produce $e^+$ in the detector, and given that it would be quite difficult
to measure the sign of the $e^\pm$ in planned detectors.  An alternate method,
to measure the asymmetry 
\beq
D = \frac{P(\nu_e \to \nu_\mu)-P(\bar\nu_e \to \bar\nu_\mu)}
{P(\nu_e \to \nu_\mu)+P(\bar\nu_e \to \bar\nu_\mu)}
\label{acpv}
\eeq
is, in principle, possible, although it is complicated by the fact, as noted
above, that $D$ is rendered nonzero by matter effects even in the absence of CP
violation (see also \cite{dgh,cpv}).  
If the solar neutrino problem is solved by 
the SMA or vacuum oscillations, CP-violation effects are not observable in the
experiments of interest here.  Indeed, even for the LMA solution, the CP
violation would be very hard to detect for path lengths larger than $\sim 3000$
km because of matter effects. 

For the Fermilab to Gran Sasso distance $L \simeq 7330$ km (or the BNL to Gran
Sasso distance $L \simeq 6560$ km), the $\Delta m^2_{sol}$ corrections are
negligible, so we can analyze the problem using fewer relevant
parameters: $\Delta m^2_{atm}$, $\theta_{13}$ and $\theta_{23}$.  We 
calculate the oscillation probabilities in long baseline experiments as a
function of $E/\Delta m^2$, rather than using a particular value for $\Delta
m^2$ or the energy.  The relevant ranges are $\Delta m^2 \sim \ {\rm few}
\times 10^{-3}$ eV$^2$ and energies $E$ of the order of tens of GeV. This way
of presenting the results can be useful in studying the optimization of the
beam energy.  We calculate the oscillation probabilities for different values
of the mixing angles $\theta_{13}$ and $\theta_{23}$ allowed by the atmospheric
neutrino data and the CHOOZ experiment.

We consider both neutrinos and antineutrinos. The matter effects reverse sign
in these two cases; for antineutrinos, $V$ in (\ref{V}) is replaced with
$(-V)$.  This implies that if $\Delta m^2$ is positive (as considered here),
one can get a resonant enhancement of the oscillations for neutrinos, while for
antineutrinos the matter effects would suppress the oscillations.  The
situation would be reversed if $\Delta m^2$ were negative.  The fact that the
matter effects are opposite in sign for neutrinos and antineutrinos is well
illustrated in Fig.\ref{fig:minos}, where both results are presented, together
with the vacuum case.

In order to study the effects at different distances, we show the same 
type of graphs for both $L=7330$ km and $L=2900$ km. For $L=2900$ km, the
probabilities can be expressed as functions of $E/\Delta m^2_{atm}$ only for
the SMA and VO solutions to the solar neutrino problem. For LMA, small 
$\Delta m^2_{sol}$ and CP violation corrections are added, as shown in 
Fig.\ref{fig:slac} and Fig.\ref{fig:cp}.

\begin{figure}[hbtp]
\begin{center}
\mbox{\epsfxsize=8truecm
\epsfysize=8.6truecm
\epsffile{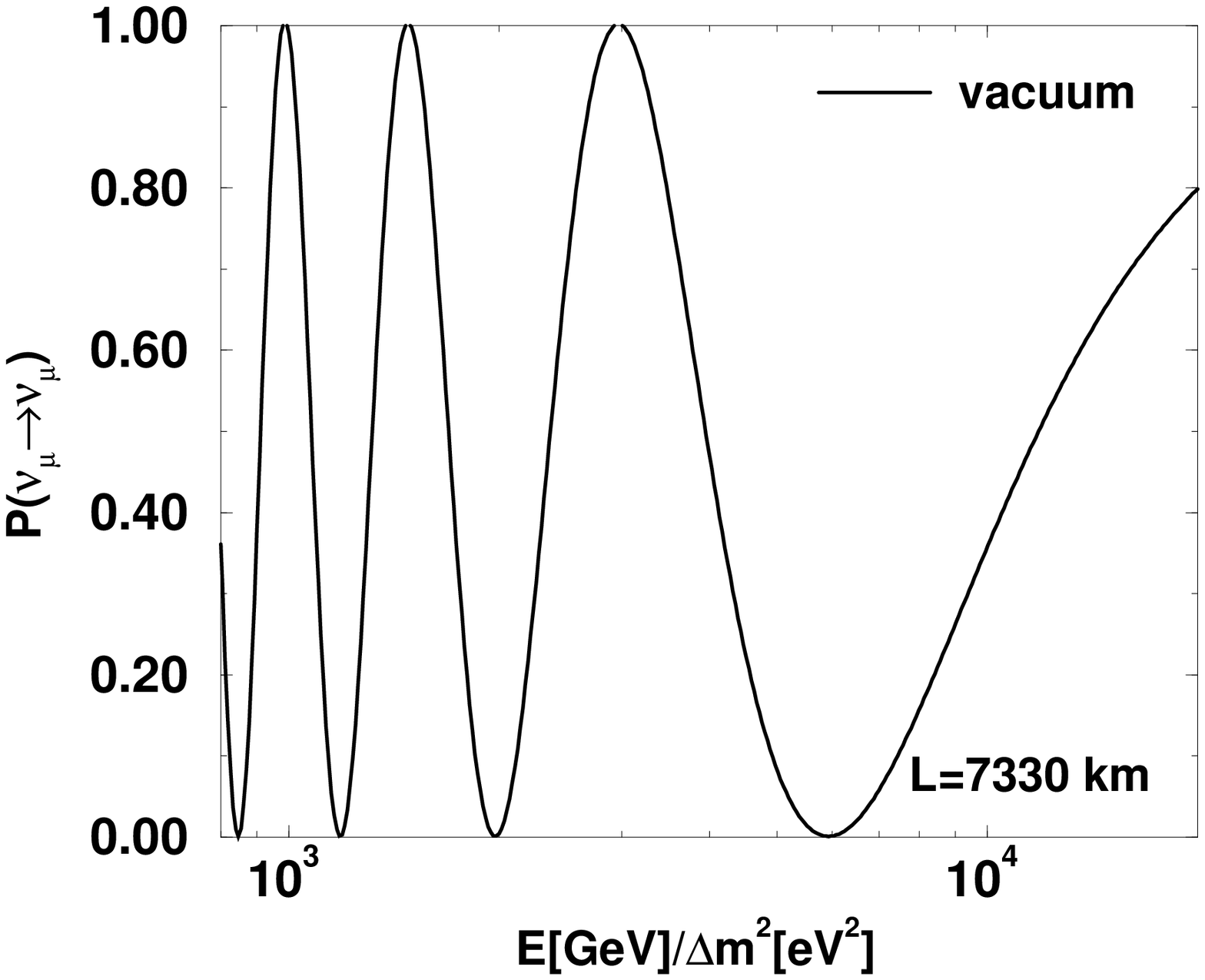}}
\mbox{\epsfxsize=8truecm
\epsfysize=8.6truecm
\epsffile{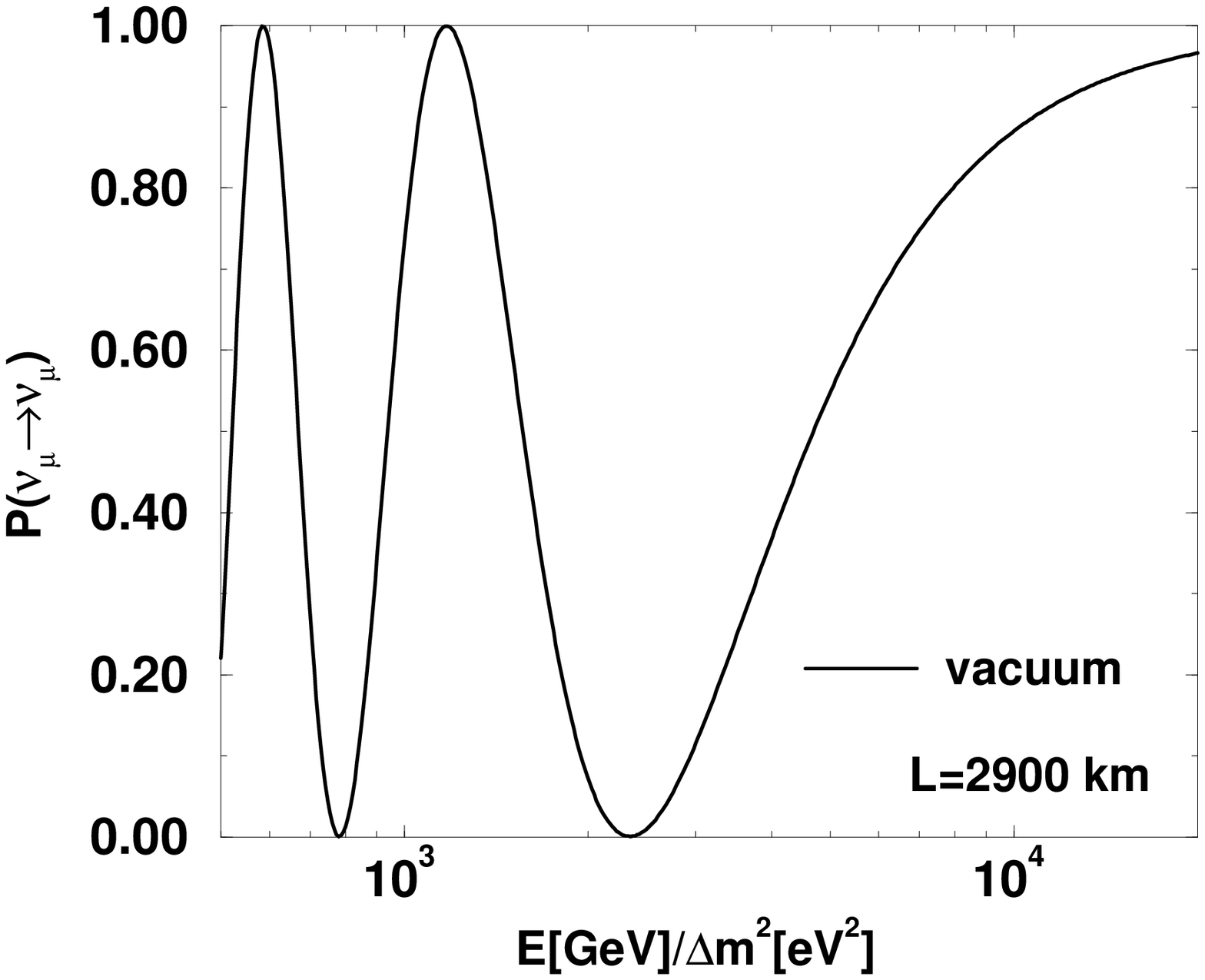}}
\end{center}
\caption{\footnotesize{Hypothetical plot of $P(\nu_\mu \to \nu_\mu)$ in 
vacuum for $L=7300$ km, $L=2900$ km with $\sin^2(2\theta_{23})=1$, for
comparison with other plots incorporating matter effects.}}
\label{fig:fgsmmv}
\end{figure}

\begin{figure}[hbtp]
\begin{center}
\mbox{\epsfxsize=8truecm
\epsfysize=8.6truecm
\epsffile{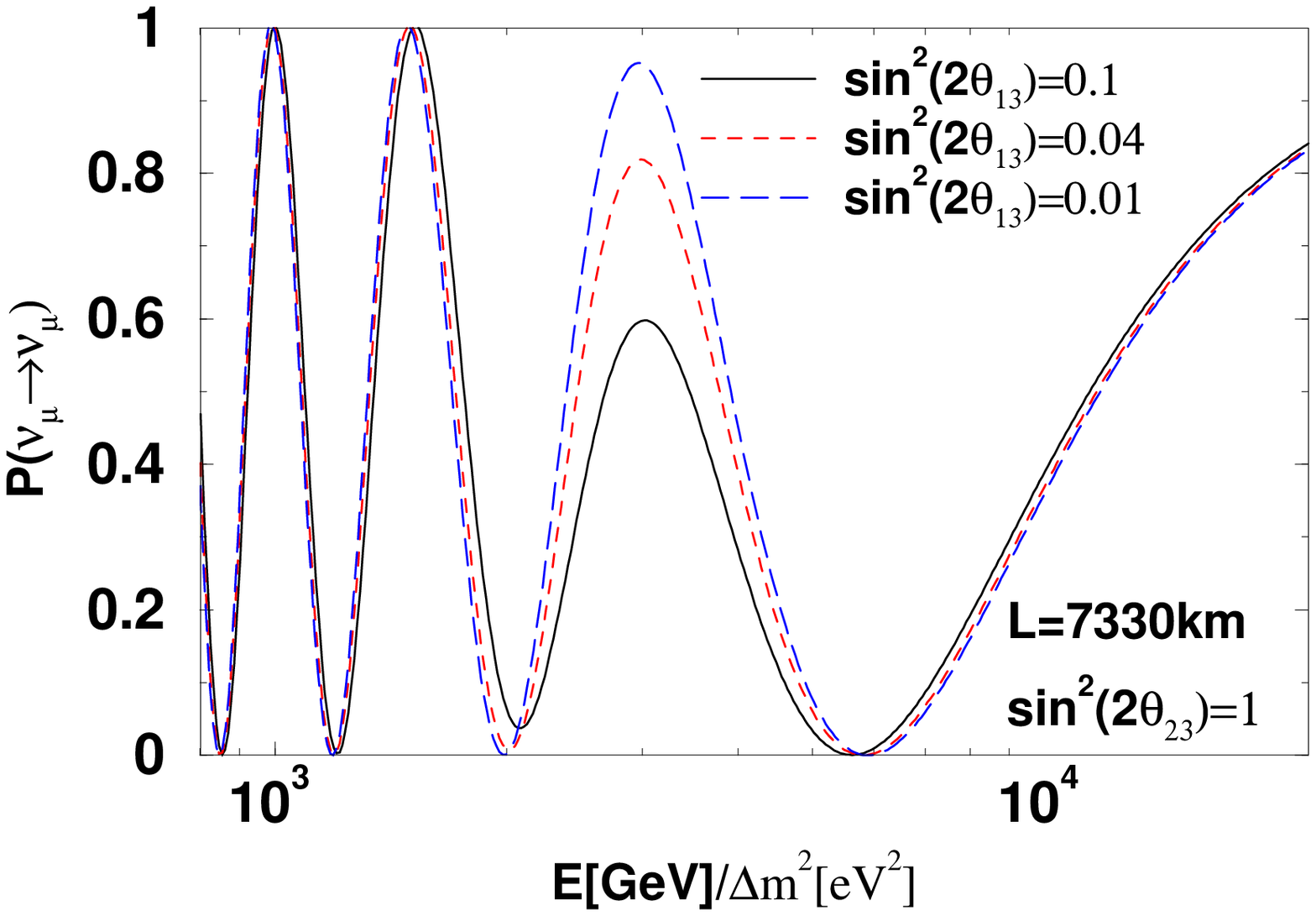}}
\mbox{\epsfxsize=8truecm
\epsfysize=8.6truecm
\epsffile{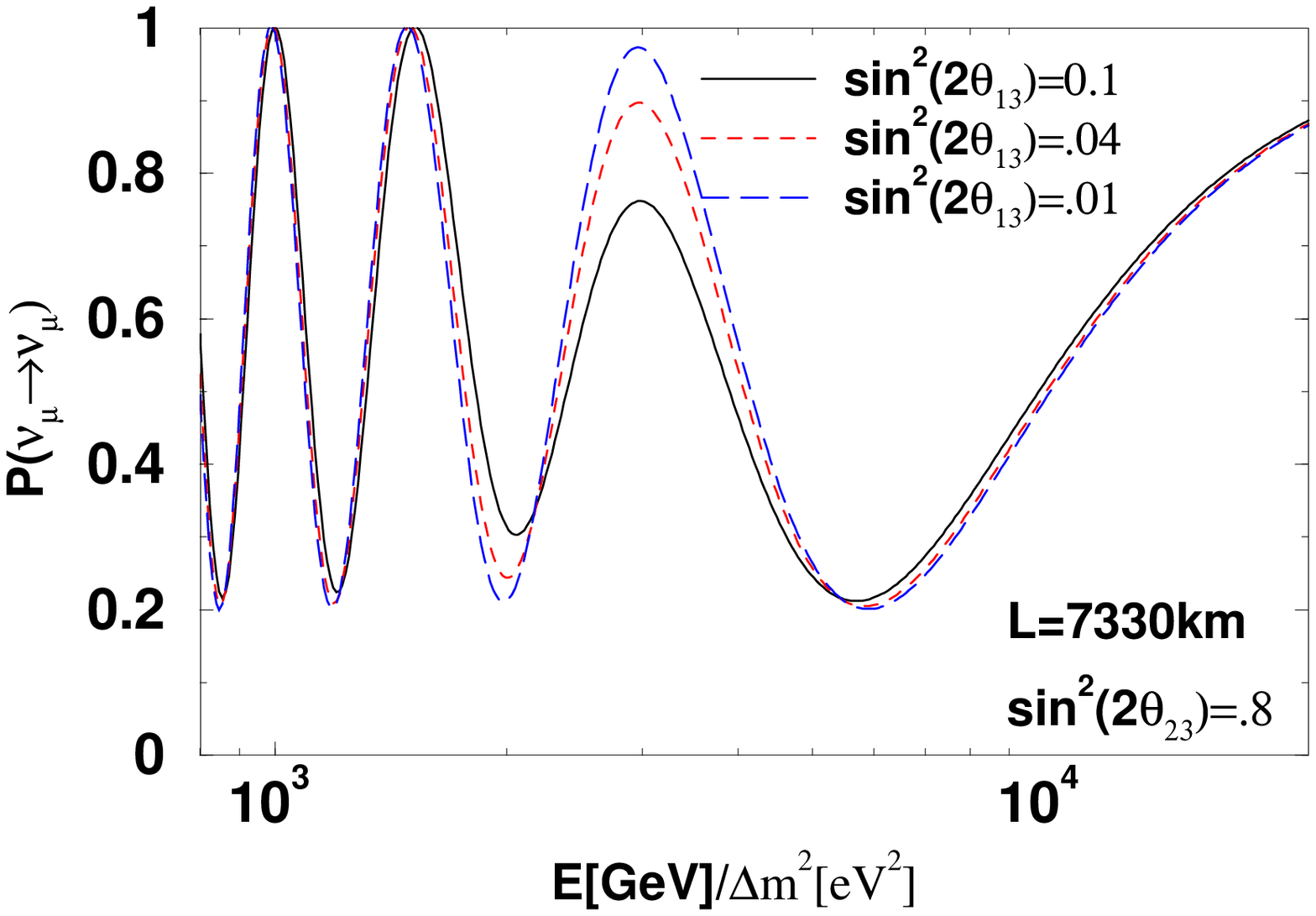}}
\end{center}
\caption{\footnotesize{$P(\nu_\mu \to \nu_\mu)$ for $L=7300$ km with 
$\sin^2(2\theta_{13})=0.1,.04,.01$ and $\sin^2(2\theta_{23})=1, 0.8$.}}
\label{fig:fgsmm}
\end{figure}

\begin{figure}[hbtp]
\begin{center}
\mbox{\epsfxsize=8truecm
\epsfysize=8.6truecm
\epsffile{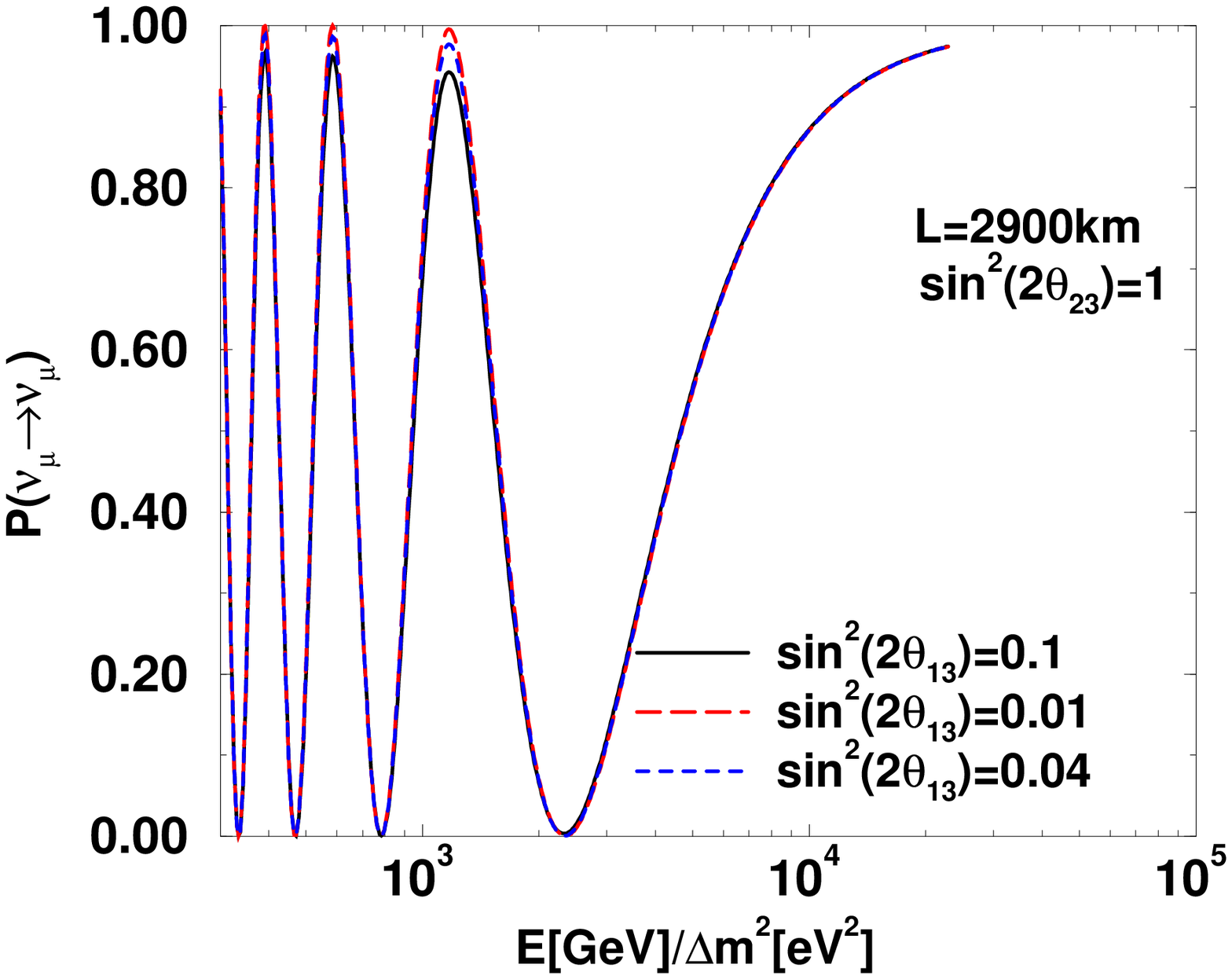}}
\mbox{\epsfxsize=8truecm
\epsfysize=8.6truecm
\epsffile{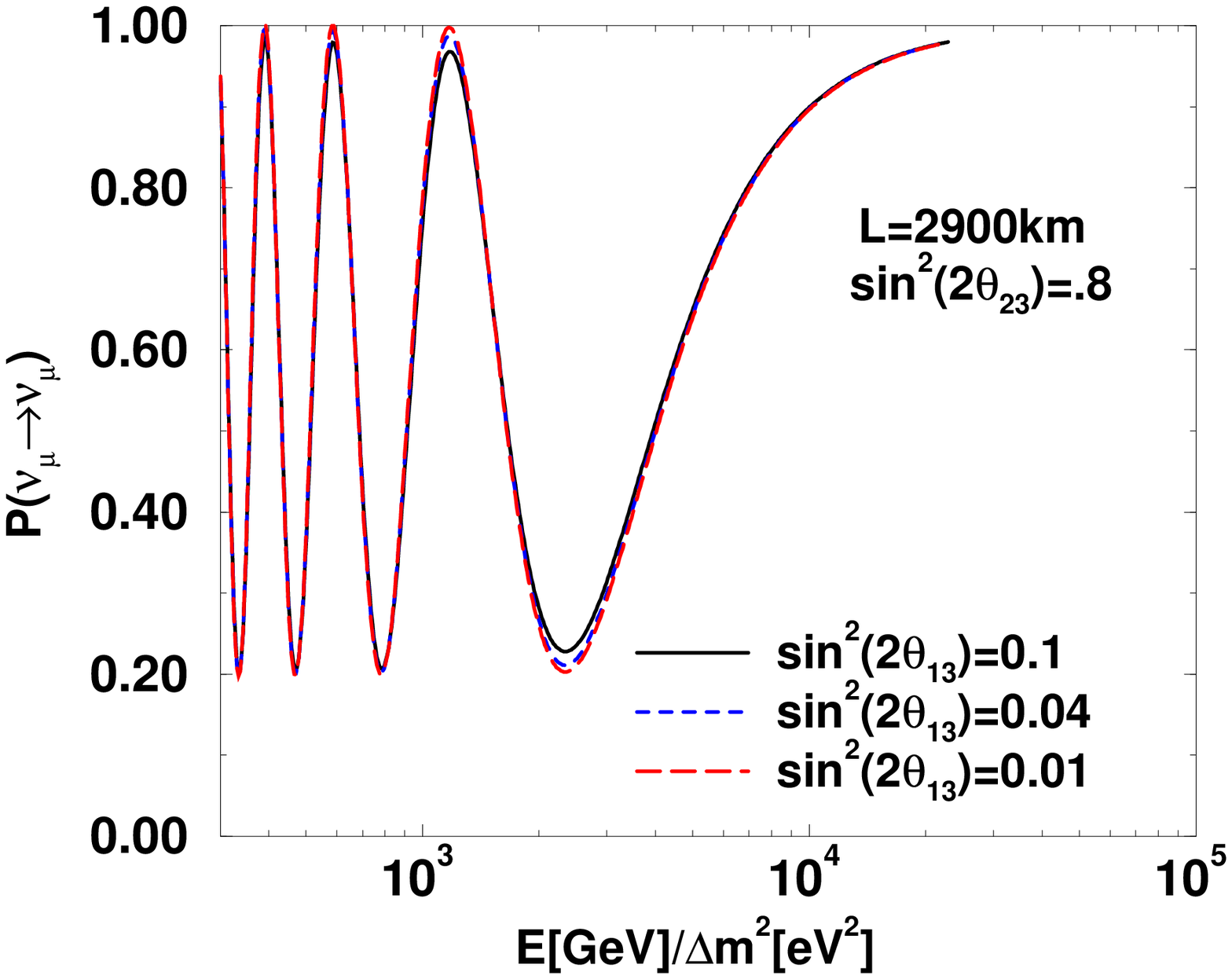}}
\end{center}
\caption{\footnotesize{Same as Fig. \ref{fig:fgsmm} for $L=2900$ km.}}
\label{fig:smm}
\end{figure}

We first study the survival probability of $\nu_\mu$. If the beam went through
vacuum, the probability of oscillation would be given by Fig.\ref{fig:fgsmmv}
for a wide range of allowed values of $\sin^2(2\theta_{13})$. In matter, this
probability becomes sensitive to all oscillation parameters for longer
baselines such as 7330 km.  In order to illustrate this, we calculate the
probability for $\sin^2(2\theta_{13})=0.1$, \ 0.04, and 0.01, and
$\sin^2(2\theta_{23})=0.8$ and $\sin^2(2\theta_{23})=1$. The results are
presented in Fig.\ref{fig:fgsmm} and Fig.\ref{fig:smm}.  Evidently, the matter
effect increases as $\theta_{13}$ increases (and vanishes if $\theta_{13}=0$).
While the shift in the positions of the maxima and minima, as functions of
$E/\Delta m^2$ are small, there is a considerable change in the maximum at
$E/\Delta m^2 \simeq 3 \times 10^3$.  This is of great interest, since the use
of a typical neutrino energy of $E \sim 10$ GeV (somewhat less than the stored
muon energy) would produce this value of $E/\Delta m^2$, given the central
value $\Delta m^2 = \Delta m_{atm}^2 \simeq 3.5 \times 10^{-3}$ eV$^2$ reported
by SuperKamiokande \cite{sk}.

\begin{figure}[hbtp]
\begin{center}
\mbox{\epsfxsize=8truecm
\epsfysize=8.6truecm
\epsffile{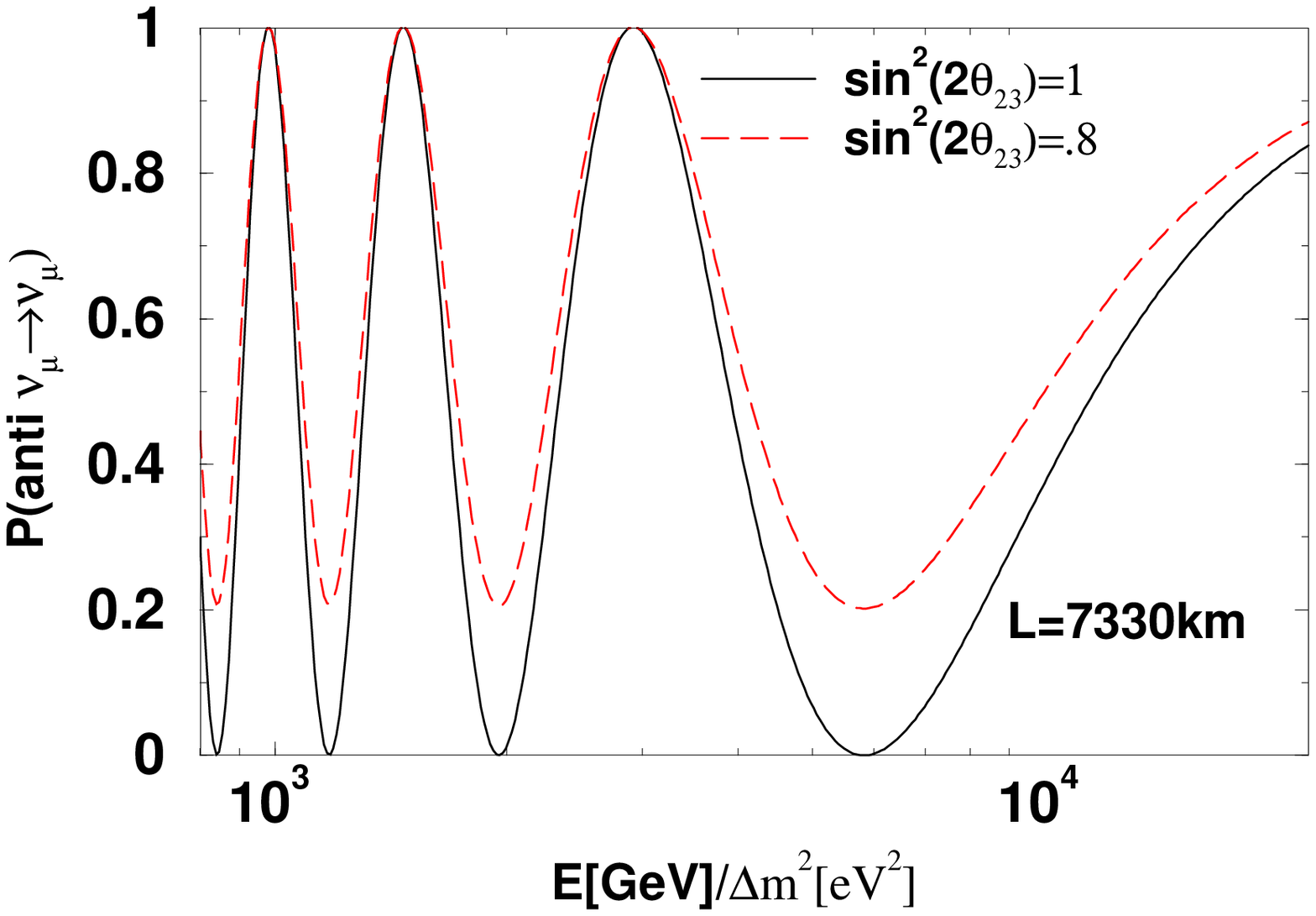}}
\mbox{\epsfxsize=8truecm
\epsfysize=8.6truecm
\epsffile{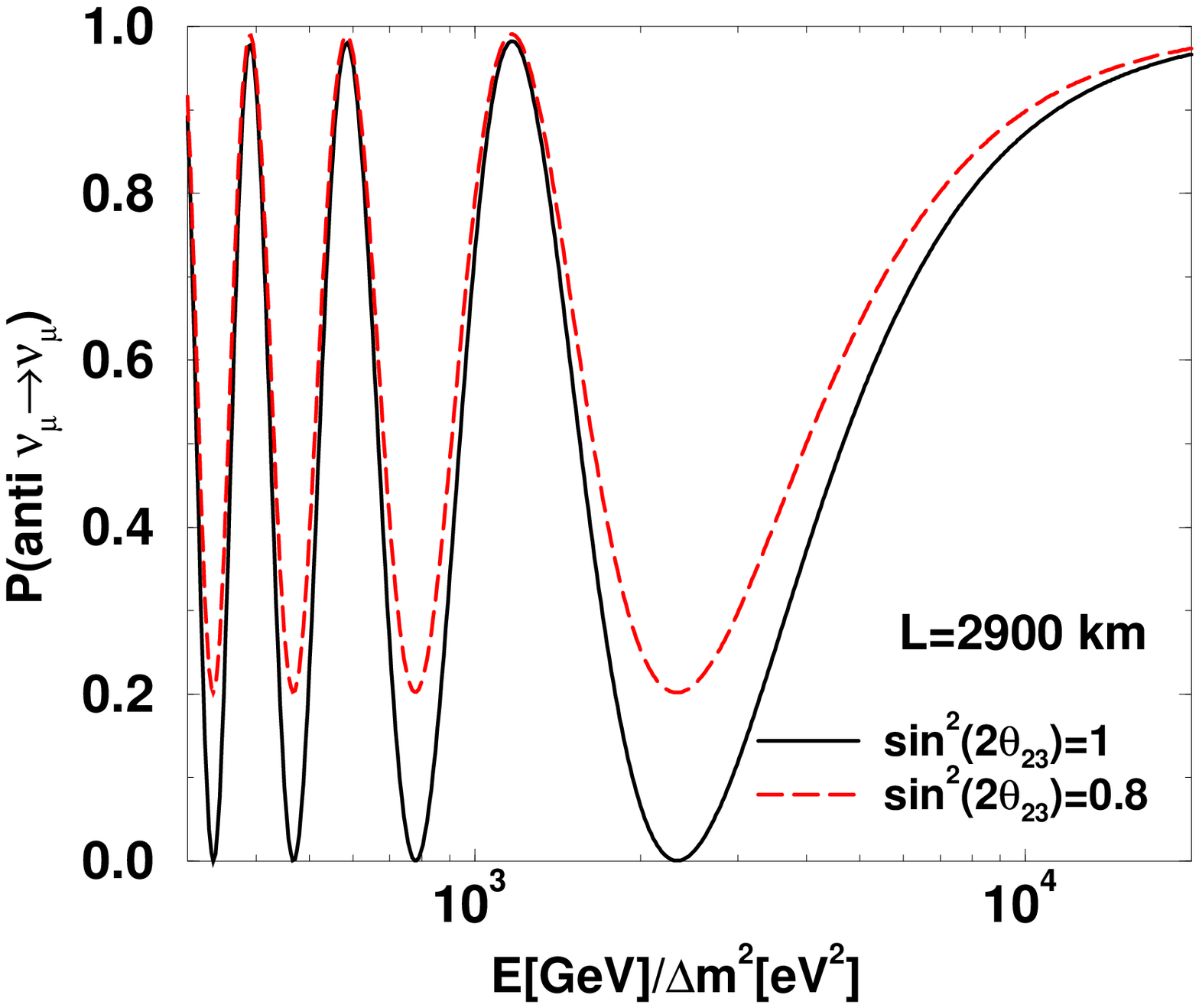}}
\end{center}
\caption{\footnotesize{$P(\bar\nu_\mu \to \bar\nu_\mu)$ for 
$L=7300$ km, $L=2900$ km with  $\sin^2(2\theta_{23})=1,0.8$ and 
$\sin^2(2\theta_{13})=0.1$.}}
\label{fig:fgsmma}
\end{figure}

We also want to compare the solution in vacuum (Fig.\ref{fig:fgsmmv}), with
the solution in matter for neutrinos (Figs. \ref{fig:fgsmm},\ref{fig:smm}) and
antineutrinos (Fig.\ref{fig:fgsmma}). For antineutrinos the $\nu_\mu$ survival
probability is not sensitive to the value of $\theta_{13}$. One can again see
the opposite effects of matter on neutrinos and antineutrinos. The difference
in the results for different mixing angles makes it possible, in principle, to
use this probability for relatively precise measurements of the oscillation
parameters.  Measuring separately the probability for $\nu_\mu$ and
$\bar\nu_\mu$ can be very useful in detecting the matter effects and using
these to constrain the relevant mixings and squared mass difference. Clearly,
if one could use two path lengths, as may be possible with a neutrino factory,
this would provide more information and constraints.

\begin{figure}[hbtp]
\begin{center}
\mbox{\epsfxsize=8truecm
\epsfysize=8.6truecm
\epsffile{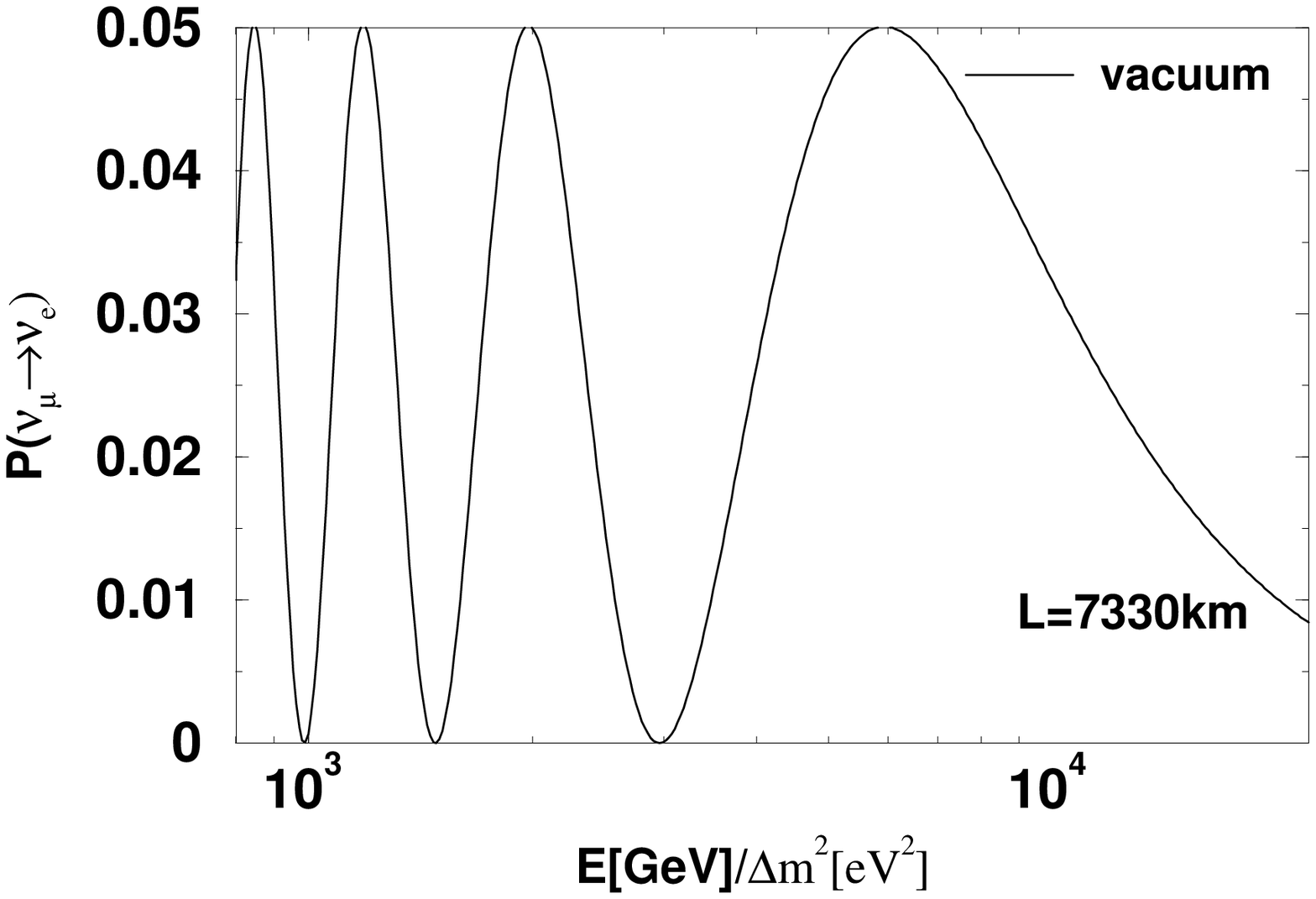}}
\mbox{\epsfxsize=8truecm
\epsfysize=8.6truecm
\epsffile{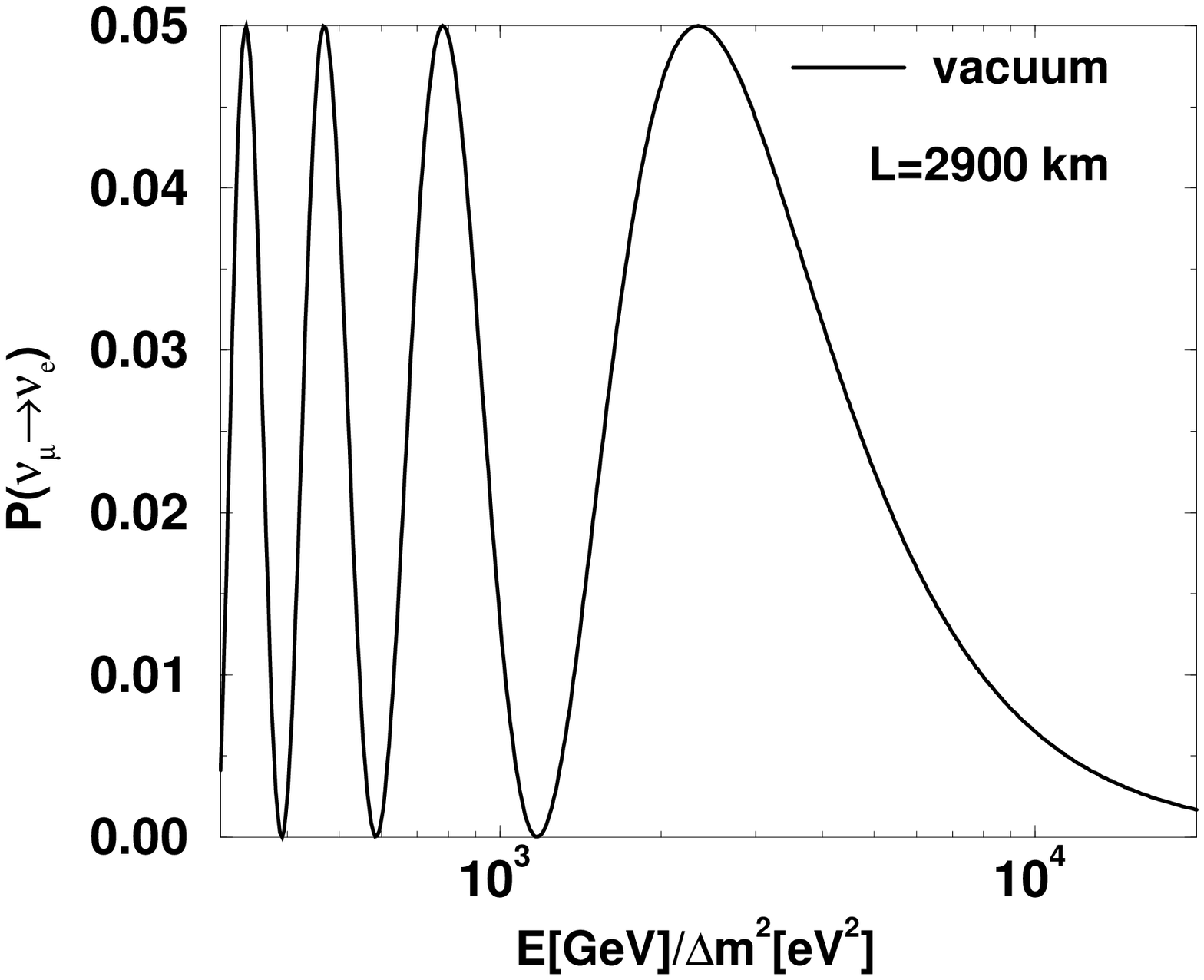}}
\end{center}
\caption{\footnotesize{Hypothetical $P(\nu_\mu\rightarrow \nu_e)$ in 
vacuum for $L=7330$ km and $L=2900$ km with $\sin^2(2\theta_{13})=0.1$
and $\sin^2(2\theta_{23})=1$ for comparison with other plots incorporating
matter effects.}}
\label{fig:fgsmev}
\end{figure}

\begin{figure}[hbtp]
\begin{center}
\mbox{\epsfxsize=8truecm
\epsfysize=8.6truecm
\epsffile{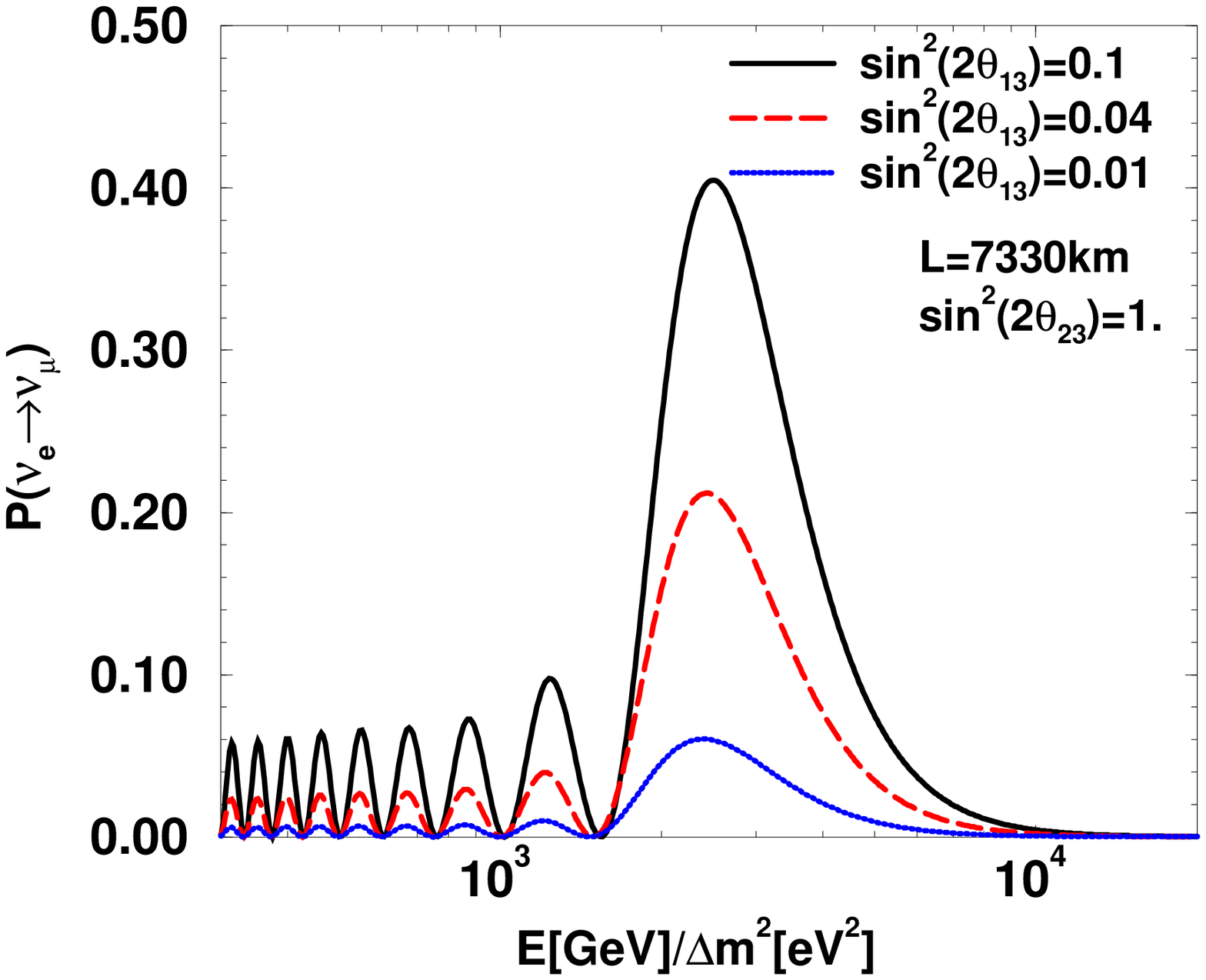}}
\mbox{\epsfxsize=8truecm
\epsfysize=8.36truecm
\epsffile{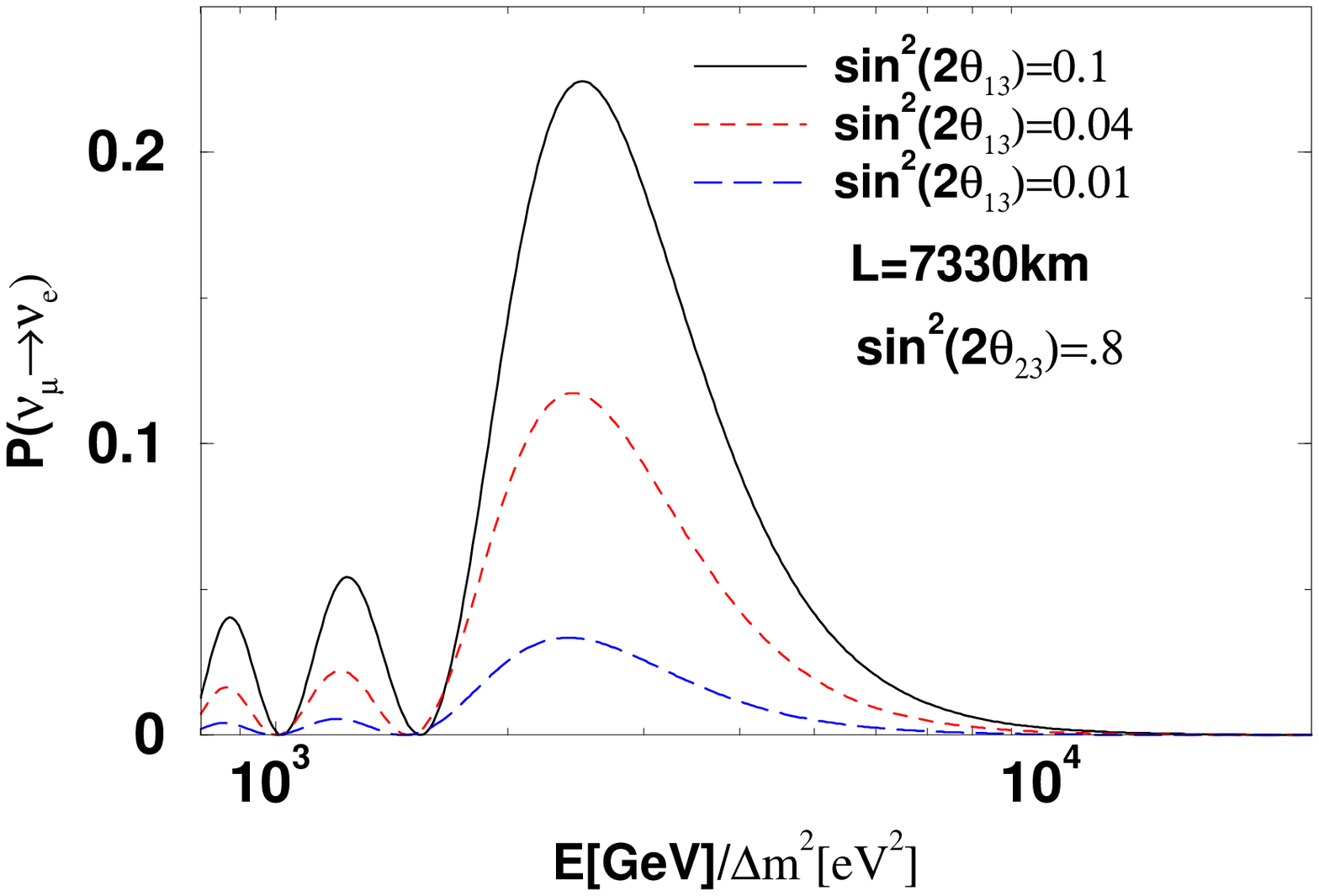}}
\end{center}
\caption{\footnotesize{$P(\nu_\mu\rightarrow \nu_e)$ 
for $L=7300$ km with $\sin^2(2\theta_{13})=0.1, 0.04, 0.01$ and 
$\sin^2(2\theta_{23})=1, 0.8$.}}
\label{fig:fgsme}
\end{figure}

\begin{figure}[hbtp]
\begin{center}
\mbox{\epsfxsize=8truecm
\epsfysize=8.6truecm
\epsffile{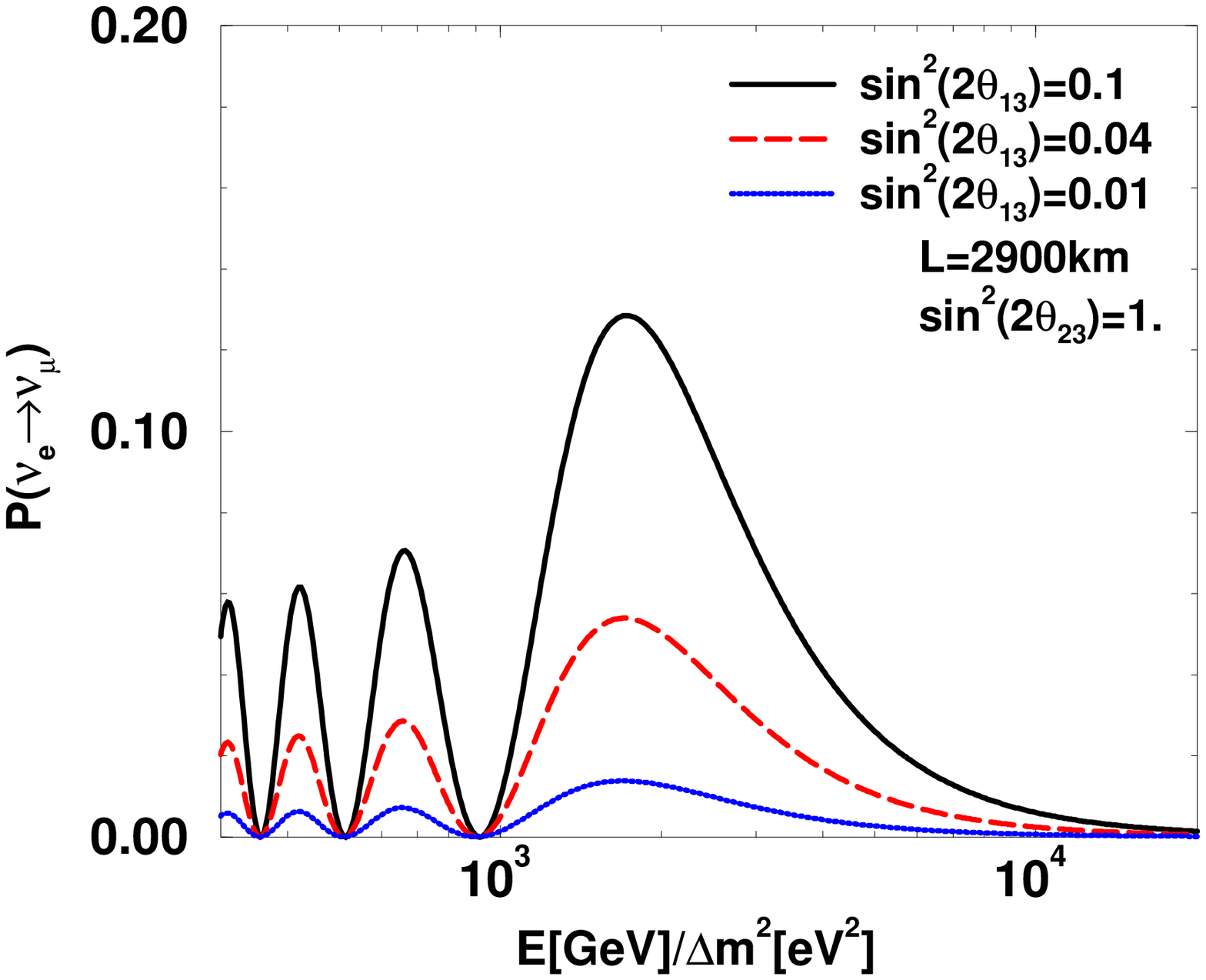}}
\mbox{\epsfxsize=8truecm
\epsfysize=8.36truecm
\epsffile{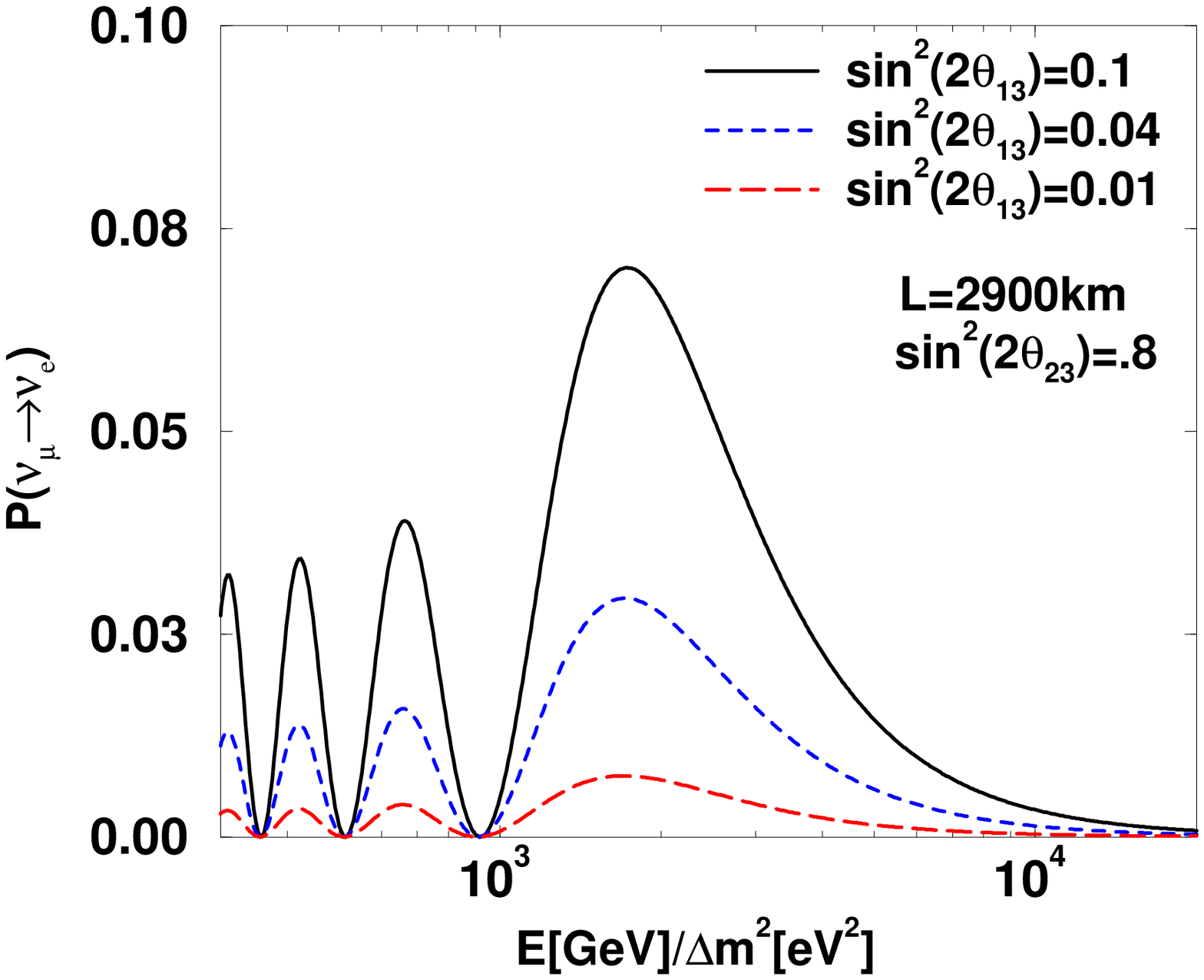}}
\end{center}
\caption{\footnotesize{Same as Fig. \ref{fig:fgsme} for $L=2900$ km.}}
\label{fig:sme}
\end{figure}

\begin{figure}[hbtp]
\begin{center}
\mbox{\epsfxsize=8truecm
\epsfysize=8.6truecm
\epsffile{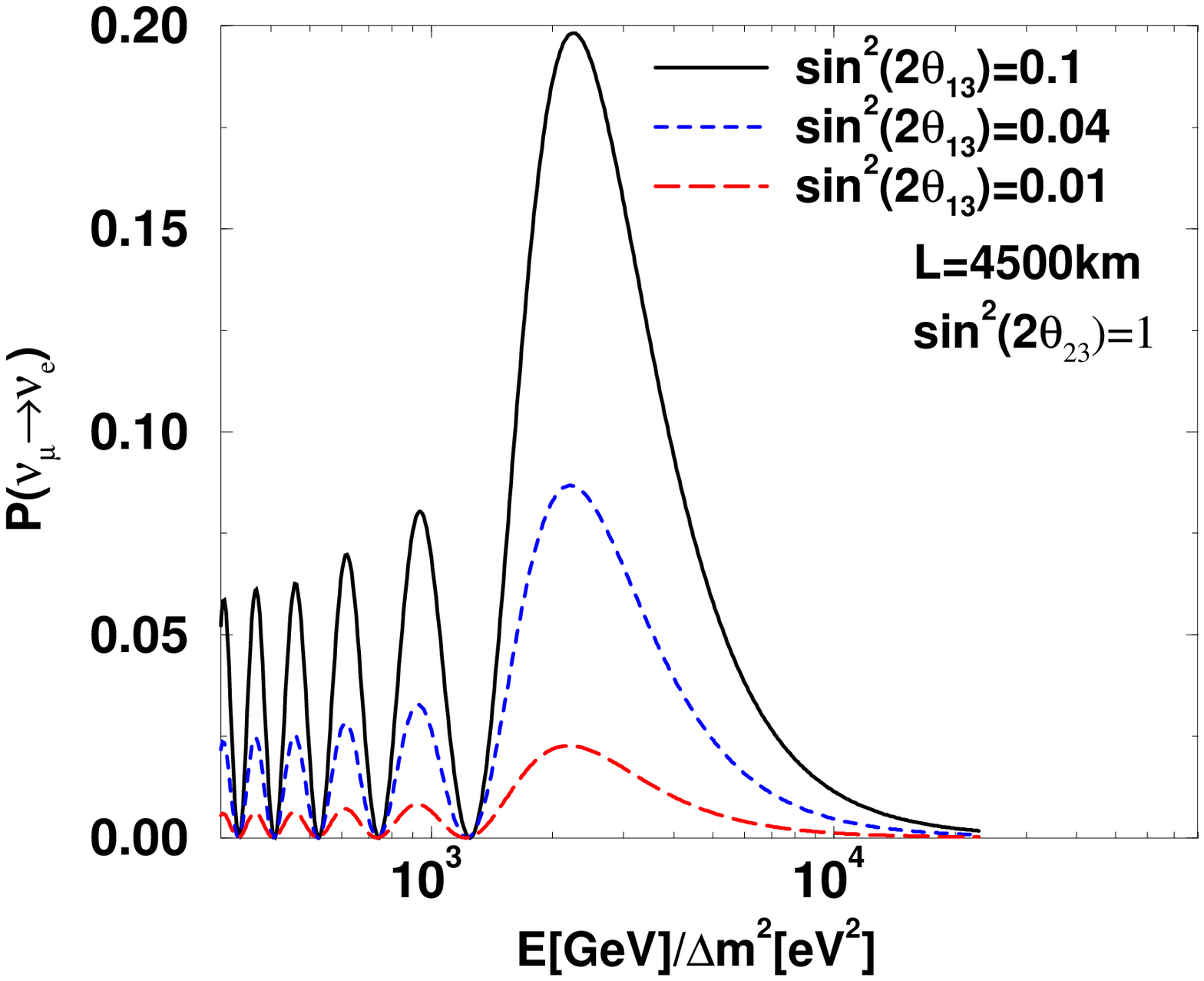}}
\mbox{\epsfxsize=8truecm
\epsfysize=8.36truecm
\epsffile{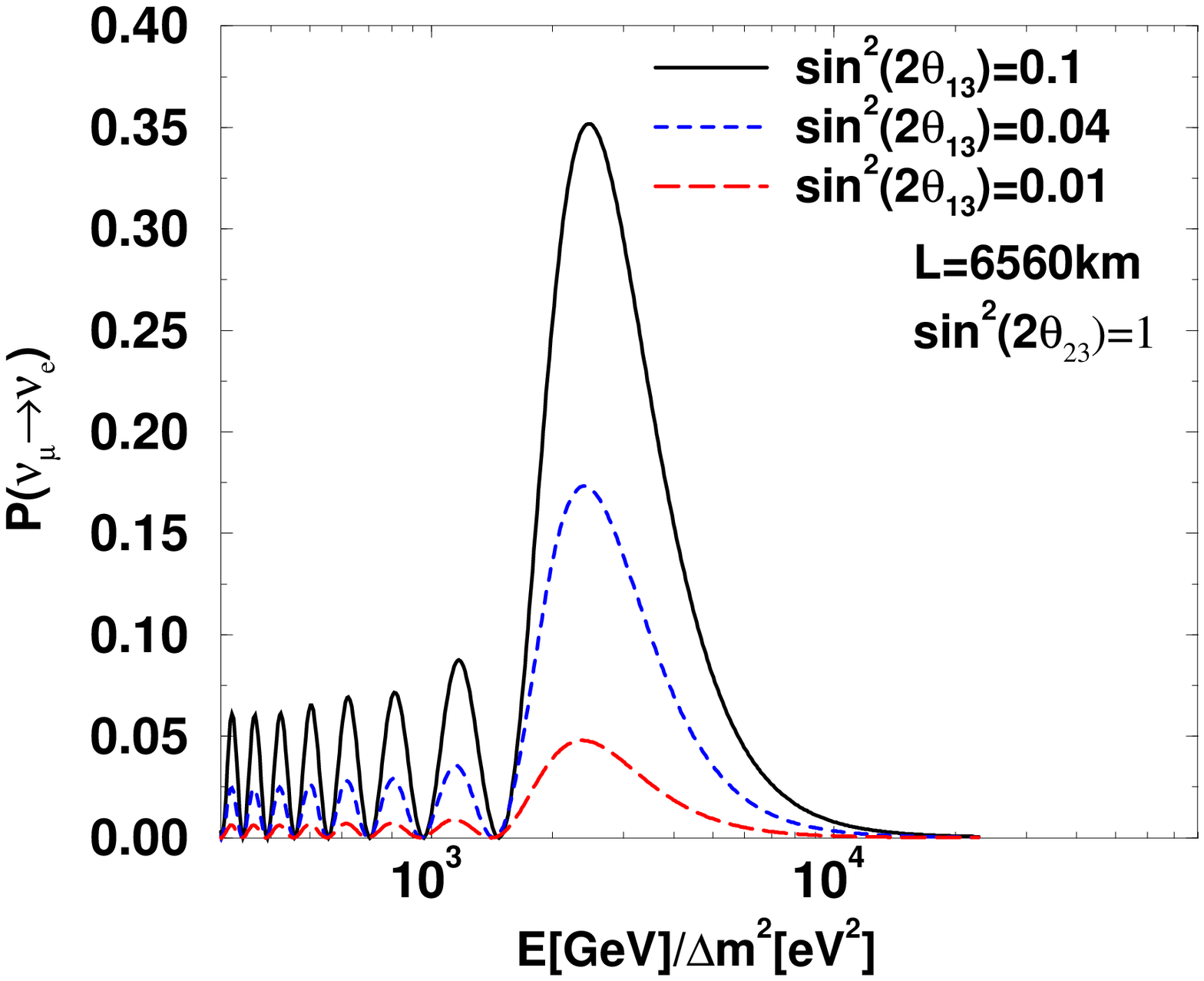}}
\end{center}
\caption{\footnotesize{$P(\nu_\mu\rightarrow \nu_e)$ 
for $L=4500$ km, $L=6560$ with $\sin^2(2\theta_{13})=0.1, 0.04, 0.01$ and
$\sin^2(2\theta_{23})=1$.}}
\label{fig:bnl}
\end{figure}

\begin{figure}[hbtp]
\begin{center}
\mbox{\epsfxsize=8truecm
\epsfysize=8.7truecm
\epsffile{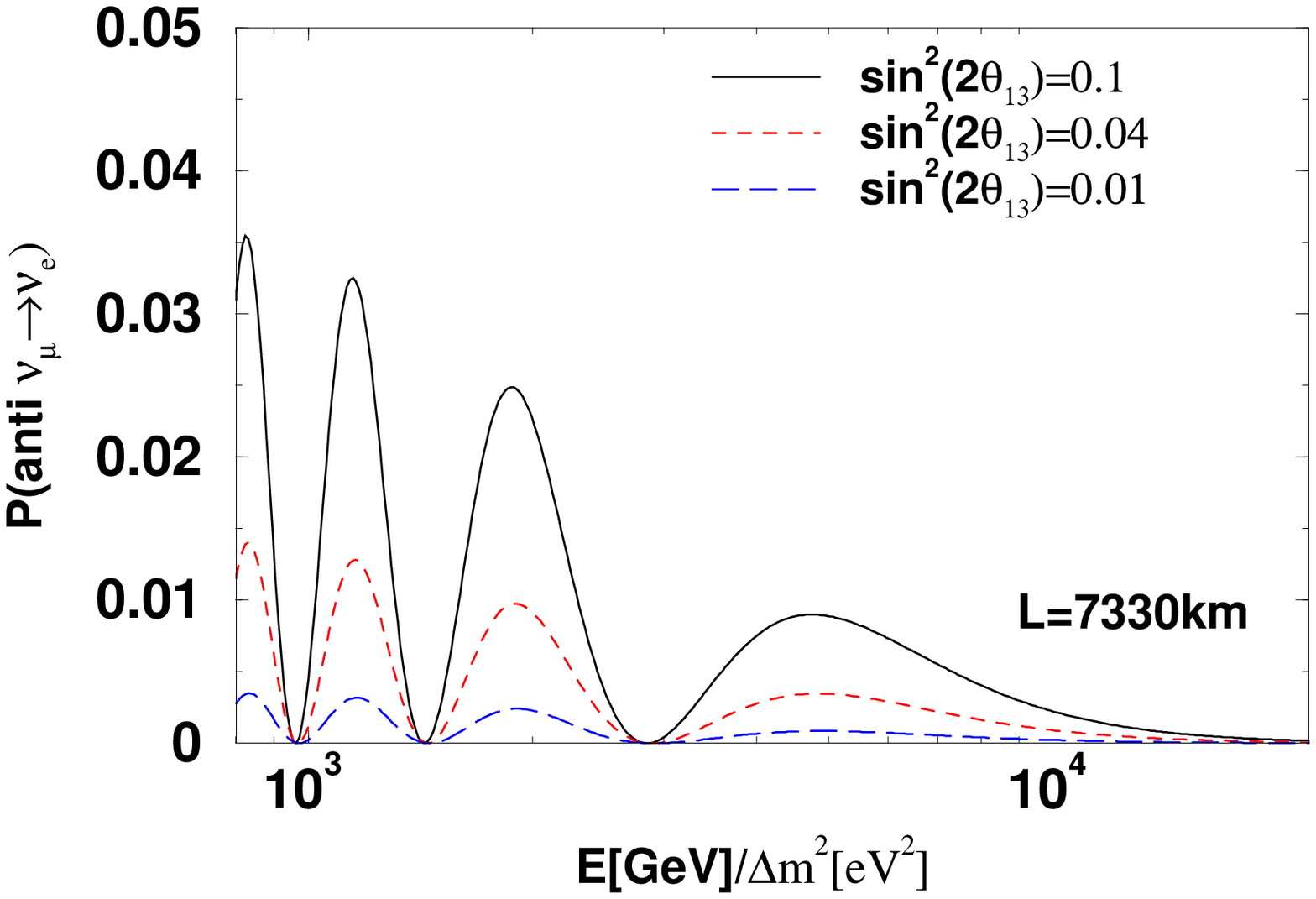}}
\mbox{\epsfxsize=8truecm
\epsfysize=8.8truecm
\epsffile{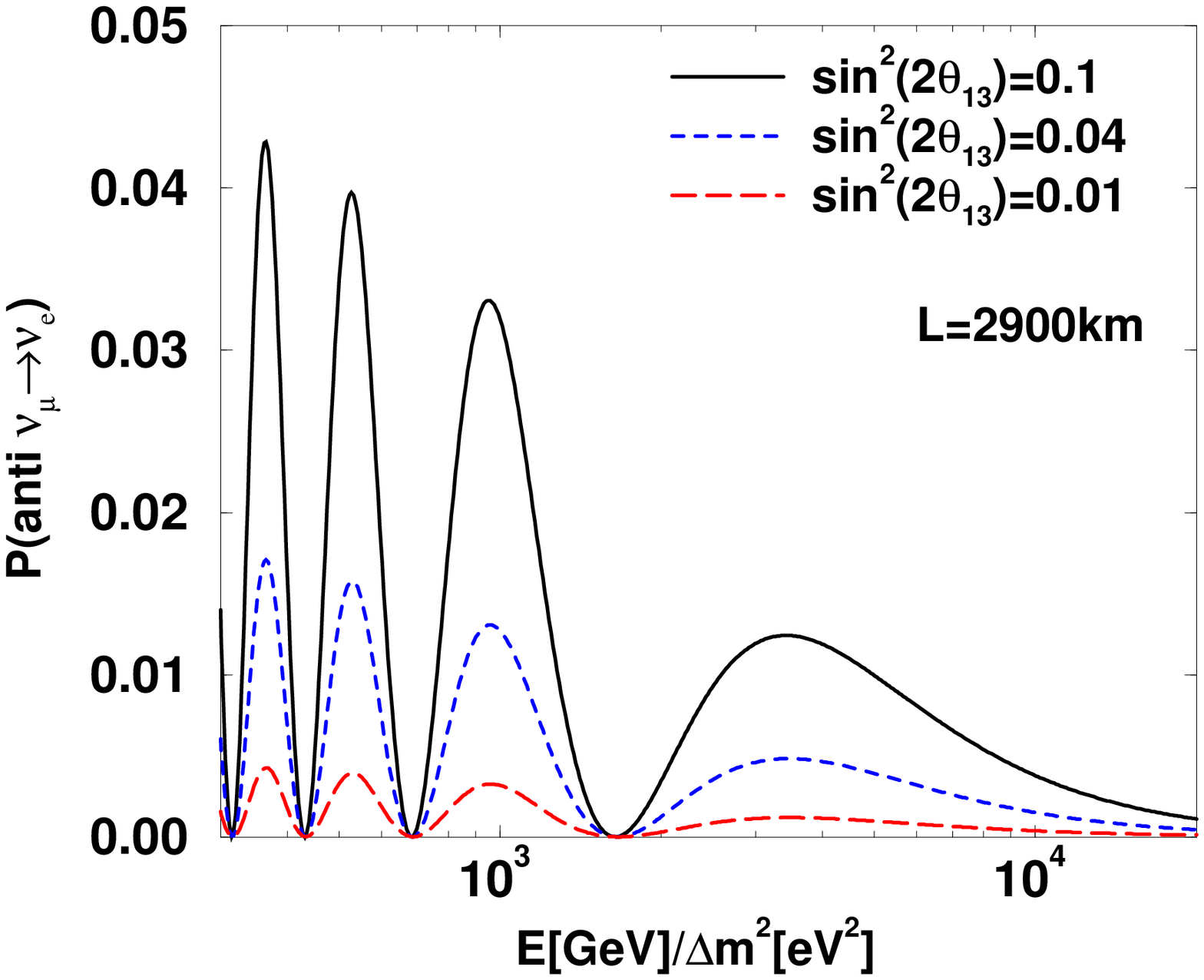}}
\end{center}
\caption{\footnotesize{$P(\bar\nu_\mu\rightarrow \bar\nu_e)$ 
for $L=7300$ km and $L=2900$ km with 
$\sin^2(2\theta_{13})=0.1, .04, .01$ and $\sin^2(2\theta_{23})=1$.}}
\label{fig:fgsmea}
\end{figure}

The relative effects of matter can be especially dramatic in the oscillation
probability $P(\nu_e \rightarrow\nu_\mu)$, since these directly involve
$\nu_e$.  Since the $\nu_e$ beam would arise from a stored $\mu^+$ beam, and
the $\bar\nu_\mu$'s from the decays of the $\mu^+$'s would produce $\mu^+$'s in
the detector, the signature for the $\nu_e \rightarrow\nu_\mu$ oscillation
would be wrong-sign muons.  As noted above, planned detectors would be capable
of searching for such wrong-sign muons.  Since this is a sub-dominant channel,
the oscillation effect is small. If the beam went through the vacuum, neither 
$P(\nu_e \to \nu_\mu)$ nor the charge conjugate, 
$P(\bar\nu_e \to \bar\nu_\mu)$, would be enhanced by the matter effect 
(see Fig. \ref{fig:fgsmev}, showing $P(\nu_\mu \rightarrow \nu_e)$, which is
equal to $P(\nu_e \rightarrow \nu_\mu)$ for the present situation where CP
violation is negligible).  Because of the matter effect however, this
probability can be strongly enhanced, as is evident in Fig.\ref{fig:fgsme} and
Fig.\ref{fig:sme}.  For $L=7330$ km, the enhancement is largest for $E/\Delta
m^2 \simeq 2.5 \times 10^3$ GeV/eV$^2$.  
This is close to the ratio that one would
get for a neutrino energy of $E \sim 10$ GeV, given the indication from the
data that $\Delta m^2_{atm} = 3.5 \times 10^{-3}$ eV$^2$. For $L=2900$ km, the
largest enhancement is obtained for $E/\Delta m^2$ a factor of 3 lower. We also
show in this case the results for the baselines corresponding to possible
BNL-SLAC and BNL-Gran Sasso distances; see Fig.\ref{fig:bnl}.  As is evident,
the matter effect can amplify $P(\nu_e \rightarrow\nu_\mu)$ and enable this
transition to be measured with good accuracy, thereby yielding very important
information on the oscillation parameters.  This probability is quite sensitive
to the value of $\theta_{13}$, so one should be able to use it for a good
determination of this angle.  This physics capability motivates careful design
studies to optimize the choice of $L$ and $E$ for this measurement.  The
sensitivity to $\Delta m^2$ is also quite strong, due to the pronounced peak
given by the matter effect in the relevant region. Note that for antineutrinos,
the oscillation is suppressed (Fig.\ref{fig:fgsmea}), so an independent
measurement of the two channels ($\nu_\mu\rightarrow \nu_e$ and
$\bar\nu_\mu\rightarrow \bar\nu_e$) would be very valuable.

\begin{figure}[hbtp]
\begin{center}
\mbox{\epsfxsize=8truecm
\epsfysize=9truecm
\epsffile{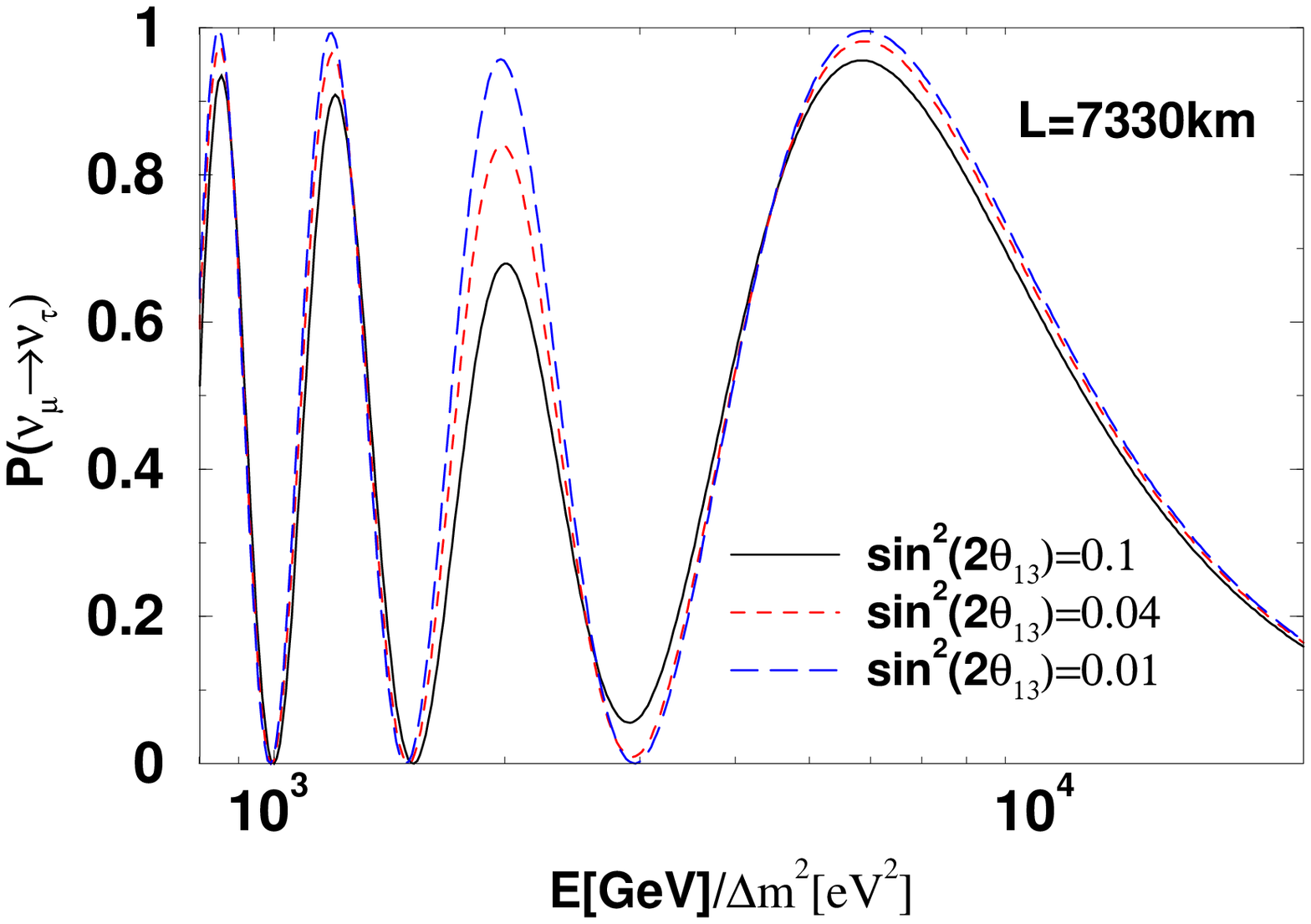}}
\mbox{\epsfxsize=8truecm
\epsfysize=9truecm
\epsffile{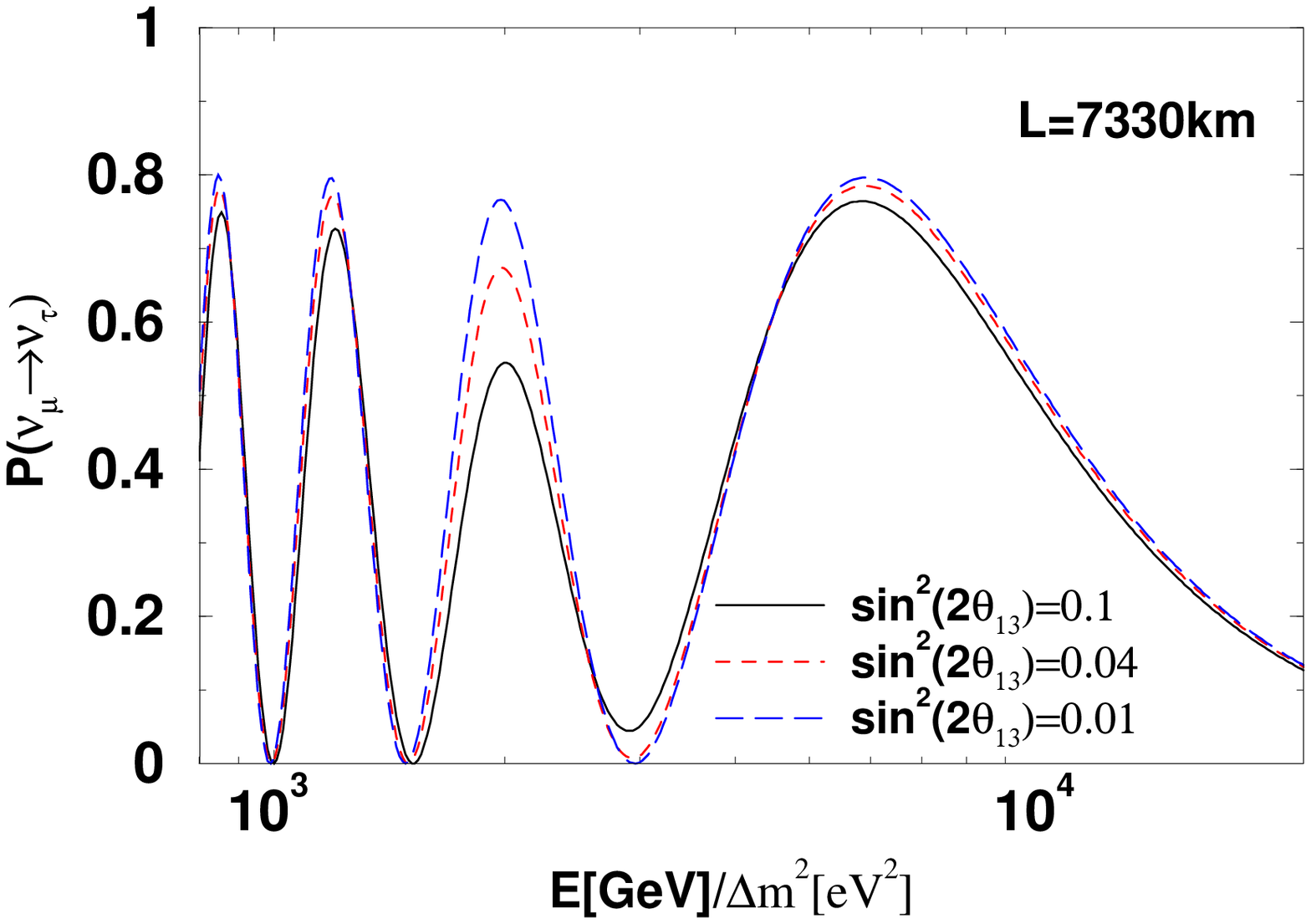}}
\end{center}
\caption{\footnotesize{$P(\nu_\mu\rightarrow \nu_\tau)$ for 
$L=7330$ km with $\sin^2(2\theta_{13})=0.1, 0.04, 0.01$ and 
$\sin^2(2\theta_{23})=1$.}}
\label{fig:fgsmt}
\end{figure}

\begin{figure}[hbtp]
\begin{center}
\mbox{\epsfxsize=8truecm
\epsfysize=9truecm
\epsffile{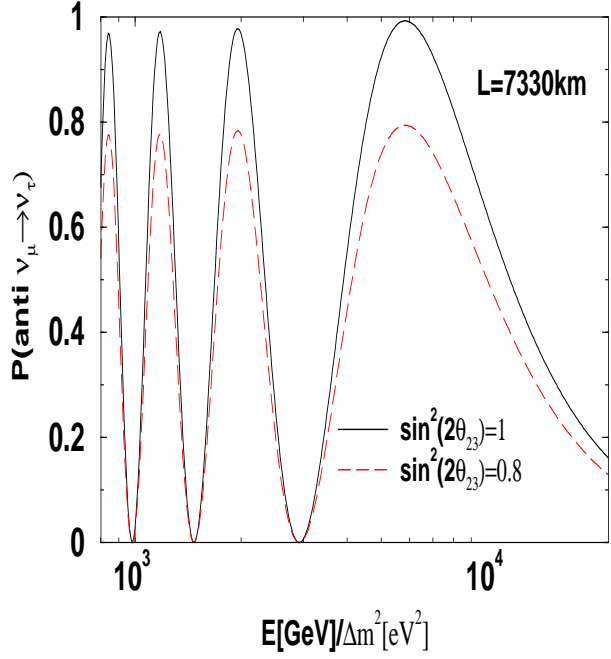}}
\end{center}
\caption{\footnotesize{$P(\bar\nu_\mu\rightarrow \bar\nu_\tau)$ 
for $L=7330$ km with $\sin^2(2\theta_{13})=0.1$ and 
$\sin^2(2\theta_{23})=1, 0.8$.}}
\label{fig:fgsmta}
\end{figure}

\begin{figure}[hbtp]
\begin{center}
\mbox{\epsfxsize=8truecm
\epsfysize=9truecm
\epsffile{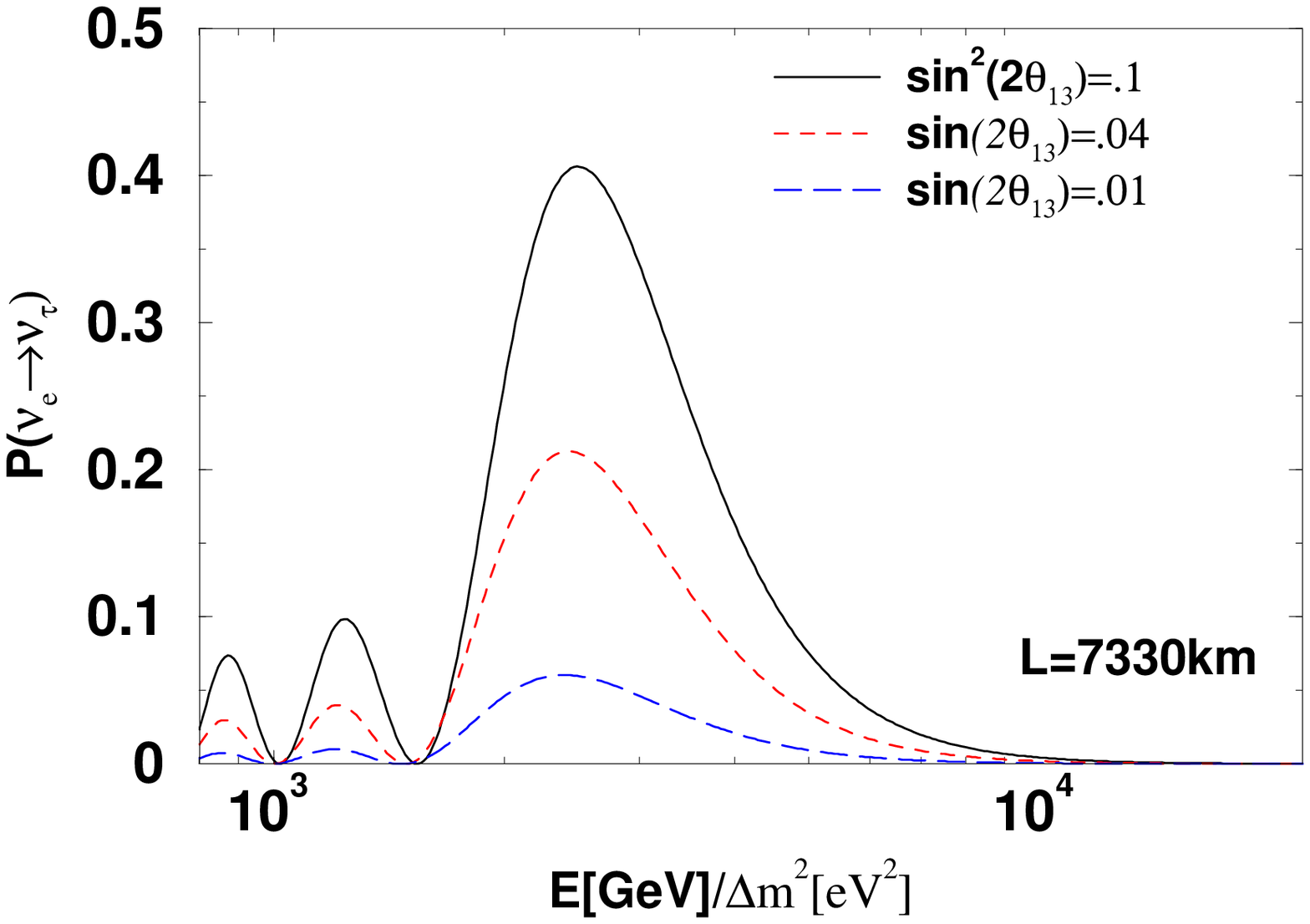}}
\mbox{\epsfxsize=8truecm
\epsfysize=9truecm
\epsffile{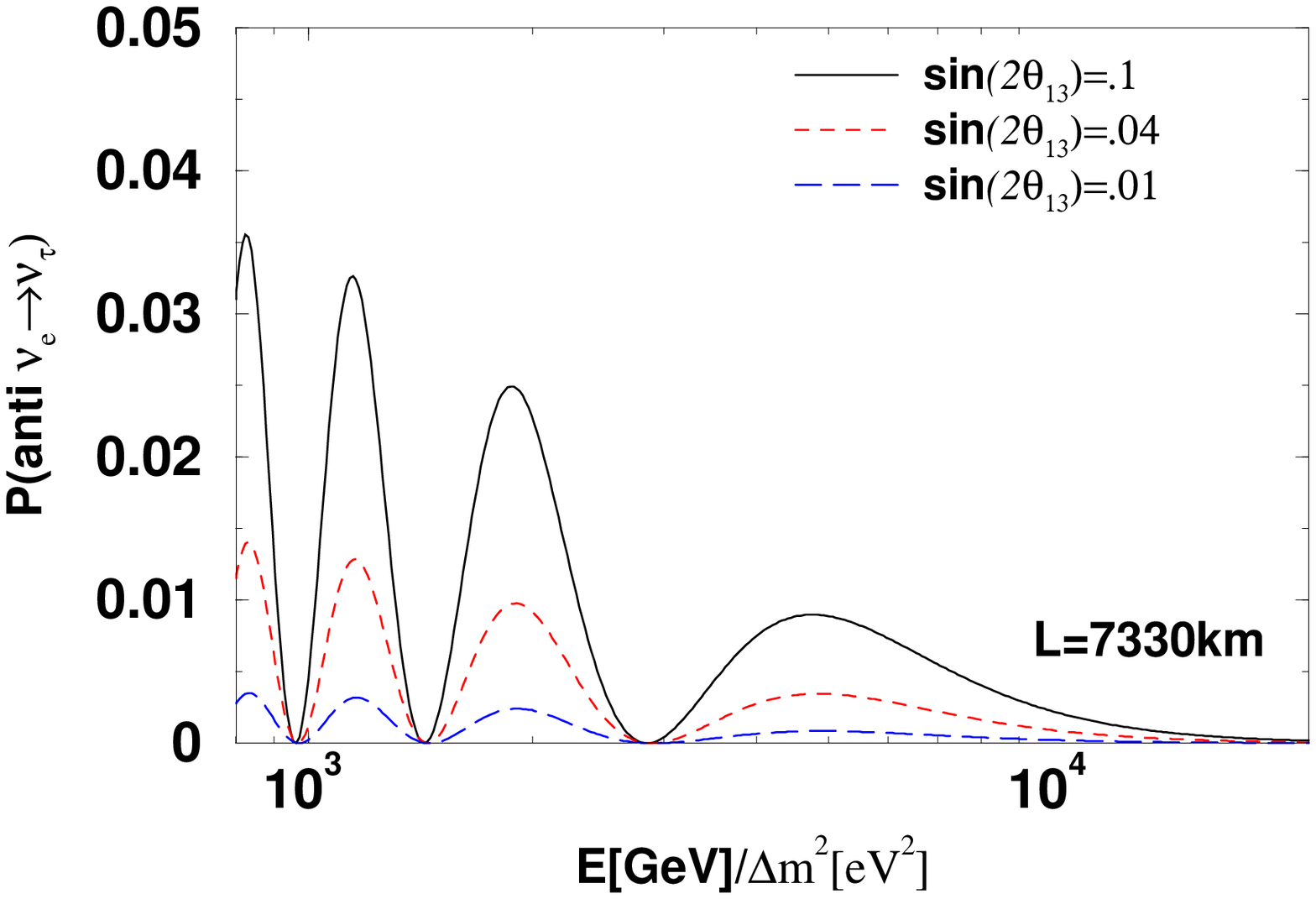}}
\end{center}
\caption{\footnotesize{$P(\nu_e\rightarrow \nu_\tau)$ and 
$P(\bar\nu_e\rightarrow\bar\nu_\tau)$ for 
$L=7330$ with $\sin^2(2\theta_{13})=0.1, 0.04, 0.01$ and 
$\sin^2(2\theta_{23})=1$.}}
\label{fig:fgset}
\end{figure}

The atmospheric neutrino data tells us that the dominant oscillation channel is
actually $\nu_\mu\rightarrow \nu_\tau$.  Consequently, it would be very useful
to measure $P(\nu_\mu\rightarrow\nu_\tau)$; this would provide further
confirmation of this oscillation and could also provide further information on
$\Delta m^2$ and $\theta_{23}$.  In addition to the MINOS experiment
\cite{anl}, the ICANOE and OPERA detectors that will operate in the CERN to
Gran Sasso neutrino beam envision $\tau$ appearance capabilities
\cite{icanoe,opera}.  Results for $P(\nu_\mu\rightarrow\nu_\tau)$ are presented
in Fig.\ref{fig:fgsmt}, and Fig.\ref{fig:fgsmta} shows
$P(\bar\nu_\mu\rightarrow\bar\nu_\tau)$.  Next, we present
$P(\nu_e\rightarrow \nu_\tau)$ in Fig.\ref{fig:fgset} and
$P(\bar\nu_e\rightarrow \bar\nu_\tau)$ in Fig.\ref{fig:fgset}.  These
calculations show that matter effects are important and enhance oscillations of
the neutrinos and suppress oscillations of antineutrinos in the relevant region
of parameters.  By combining results from different types of measurements, in
different channels of oscillations, the allowed parameter space can be strongly
constrained, leading to precise measurements of all mixings and $\Delta m^2$.

For a baseline over 9000 km, as would be the case for an experiment from
Fermilab to SuperKamiokande, the main features discussed above remain
true. Matter effects are significant for the oscillation of neutrinos for
$\Delta m^2$ in the region suggested by the atmospheric data and energies of
the order of 10 GeV . Due to the matter effects, oscillations probabilities 
become very sensitive to $\theta_{13}$. Matter effects also improve
the sensitivity to $\Delta m^2$.  Since matter effects for antineutrinos are
opposite to those for neutrinos, independent measurements of neutrino and
antineutrino oscillations would give a precise measure of the matter effects
and, consequently, of the parameters relevant to the oscillations.  Due to the
longer path through the Earth with bigger density, the matter effects can
become even more dramatic.  However, the statistics of the experiment would be
limited by the lower neutrino flux at larger distances, and a careful
study is necessary in order to choose optimal values of $L$ and $E$. 

\section{Summary}

To summarize, in planning for very long baseline neutrino oscillation
experiments, it is important to take into account matter effects.  These
effects are significant for the range of neutrino energies $E$ of order 10's of
GeV that are planned for these experiments, given the density of the Earth and
the value of $\Delta m_{atm}^2 \sim 3 \times 10^{-3}$ eV$^2$ indicated by
current atmospheric neutrino data.  We have performed a study of these
including realistic density profiles in the earth.  Matter effects can be
useful in amplifying neutrino oscillation signals and helping one to obtain
measurements of mixing parameters and the magnitude and sign of $\Delta
m^2_{atm}$.
 
\vspace{6mm}

{\bf Acknowledgments} 

\vspace{3mm} 

We thank Bob Bernstein and Debbie Harris for helpful comments and have
benefitted from participation in the ongoing working groups studying the design
and physics capabilities of storage rings as neutrino factories
\cite{web,bnlbook}.  The research of R. S. was supported in part at Stony Brook
by the U. S. NSF grant PHY-97-22101 and at Brookhaven by the U.S. DOE contract
DE-AC02-98CH10886.\footnote{\footnotesize{Accordingly, the U.S. government
retains a non-exclusive royalty-free license to publish or reproduce the
published form of this contribution or to allow others to do so for
U.S. government purposes.}}

\newpage

\end{document}